\documentclass{aa}  

\usepackage{graphicx}

\usepackage{txfonts}

\usepackage{hyperref}
\usepackage{threeparttablex}
\usepackage{xcolor}
\usepackage{amsmath}
\usepackage{caption}

\begin{document} 

   \title{From simulations to observations of cool plasma: What we can learn from synthetic spectra}

   \author{V. Jer\v{c}i\'{c}\inst{1}
          \and
          T. A. Kucera\inst{1}
          \and
          A. G. M. Pietrow\inst{2}
          \and
          P. Antolin\inst{3}
          \and
          J. M. Jenkins\inst{4}
          }

   \institute{NASA Goddard Space Flight Center, Greenbelt, MD, USA\\
              \email{veronika.jercic@nasa.gov}
        \and
             Leibniz-Institut für Astrophysik Potsdam (AIP), An der Sternwarte 16, 14482 Potsdam, Germany
         \and
             School of Engineering, Physics and Mathematics, Northumbria University, NE1 8ST, Newcastle upon Tyne, UK
         \and
             European Space Agency (ESA), European Space Astronomy Centre (ESAC), Camino Bajo del Castillo s/n, 28692 Villanueva de la Cañada, Madrid, Spain
             }

   \date{Received ??; accepted ??}

  \abstract
 
   {Cool plasma of coronal rain or prominences and filaments is ubiquitous in the solar corona. Spectroscopic observations of it have been gathered for decades; however, they have proven difficult to interpret.  }
   {Improved diagnostics would allow us to better disentangle the complex atmosphere at the source of these spectra and directly relate specific observables to the conditions in the atmosphere. We tried to achieve this by exploring the synthetic spectra of Mg~II~h\&k lines in detail and analysing how well a non-adiabatic magnetohydrodynamic (MHD) simulation matches observations when comparing the synthetic with the observed spectra. }
   {A  Python implemented, non-local thermodynamic equilibrium (non-LTE) radiative-transfer framework, \texttt{Promweaver}, was recently developed specifically for condensations in the corona. We applied \texttt{Promweaver} to create synthetic spectra from a 2.5D MHD simulation created with \texttt{MPI-AMRVAC}. The simulation shows a dynamic system of numerous cold, dense thread-like structures that we compare to observations of the \textit{Interface Region Imaging Spectrograph} (IRIS), with a focus on flare-driven coronal rain. }
   {We give a detailed description of the simulation from the aspect of the synthetic spectra. We show how and why the particular spectra form. A single thread already demonstrates great complexity that can influence the shape of the observed spectral line. We further used the quartile analysis to systematically explore the evolution of observables in the observed and the synthetic spectra. This gives an overview of how the parameters change in time and that the changes seen in the simulation fit with the changes seen in observations. The results show that the simulation, in some aspects, matches particularly well to the observations of flare-driven coronal rain. Furthermore, we explored relations of parameters of the simulation and the observables. We find significant correlation between intensity and density, but not with temperature (in the studied temperature range of 5 to 20\,kK). As a result, there is also a correlation of intensity and pressure. This relation allows us to further relate intensity and the vertical component of the Lorentz force. All of this brings us a step closer to a better understanding of the connection between the observed spectra and the atmosphere from which it originates.}
   
    {}
    

   \maketitle
   \nolinenumbers

\section{Introduction}

   Although the temperature in the solar corona is generally thought to be $\ga10^6$~K, it also contains regions of significantly cooler plasma. Prominences, also known as filaments (if seen on the solar disc rather than off the limb), are plasma structures two orders of magnitude colder and denser than the hot and tenuous solar corona in which they primarily reside. They can be sustained for a relatively long time (days, weeks) due to the magnetic field supporting them \citep[although individual blobs and threads can change on much shorter scales of tens of minutes;][]{Lin2005}. They appear in a variety of shapes, sizes, and dynamics. For more details we refer the reader to reviews such as \cite{Mackay2010}, \cite{Labrosse2010}, and \cite{Vial_Engvold2015}. Another form of cold and dense plasma in the solar corona that has recently been gaining a lot of attention is coronal rain \citep{Antolin2020}. Coronal rain does not persist for a long time in the corona; individual blobs of cold and dense material fall to lower layers of the atmosphere relatively quickly \citep[in tens of minutes; see][]{Antolin2015}. 
   
   Due to their particular composition (cold, partially ionised plasma) prominences and filaments look very different in observations. In the 'cold' lines, such as H$\alpha$ and Mg~II  (in which we often observe them), prominences are seen above the limb in emission. As coronal plasma does not gain sufficient opacity to be visible in these lines, we clearly see only the prominence as it absorbs and then scatters the incident radiation coming from the solar disc below. On the other hand, filaments are dark structures that are always observed on disc in absorption in H$\alpha$. They also absorb and scatter the emission, but their scattering along the LOS is much fainter than what the chromosphere emits in these lines that we do not actually see these structures \citep{Heinzel_chapter2015}, i.e. they outline the chromosphere whose light they block. Coronal rain, although often observed above the limb, shares the same absorption and emission characteristics and hence has the same appearance when observed on disc \citep[for observations of coronal rain on disc, see][]{Antolin2012, Pietrow2024_flares, De_Wilde2025}. 
      
   Even though predominantly in the corona, prominences and coronal rain connect all the layers of the solar atmosphere, making them a powerful tool for probing the balance and energy transfer between the different layers. Despite having much larger densities than the corona, both prominences and coronal rain are not able to maintain a local thermodynamic equilibrium \citep{Labrosse2010, Heinzel2025}. On average, these condensations have temperatures of the order of 10$^4$\,K or lower \citep{Parenti2014, Antolin2015}, and at these temperatures both ions and neutrals are present \citep{Ballester2018,Parenti2024,Jercic2025}. Due to their complex thermal structure, this cool plasma has been posing a challenge for scientists for decades. Understanding and recreating how light is absorbed and emitted is no small feat when it comes to non-local thermodynamic equilibrium (non-LTE).

   The Mg~II resonance lines in particular have strong diagnostic capabilities, and they are often used for probing the chromosphere but equally important for probing similarly cold plasma in the corona \citep{Heinzel2014, Heinzel2015_letter, Vial2016, Vial2019, Schwartz2024}. While Mg~II~h and k are strong resonance lines, there are also weaker triplet lines---one located in the blue wing of the k line and two in its red wing---between h and k \citep[with vacuum wavelengths of 2791.60, 2798.75, and 2798.82\,\AA, respectively,][]{Pereira2015,Kerr2015}. The Mg II resonance lines originate from a transition between the 3s ground level and the 3p level. The weaker triplet lines are further formed from the 3p level populating the higher 3d one. Thus, the formation of the triplet lines is dependent on formation of the h and k lines \citep{Levens2019}.
   
   \cite{Lemaire_Gouttebroze1983} modelled Mg~II line formation within prominences. They used a model with up to 13 levels to analyse how higher levels influence the resonance-line profiles.  \cite{Paletou1993} modelled Mg~II~h\&k using a 2D isothermal, isobaric slab with partial frequency redistribution \citep[PRD; frequencies of absorbed and emitted photons are correlated; see also][]{Vial_Engvold2015} and irradiated by an incident light based on observational values (OSO-8). With the launch of the \textit{Interface Region Imaging Spectrograph} \citep[IRIS;][]{DePontieu2014} in 2013 also came a vast amount of observational data of Mg~II~h and k lines in the cool plasmas of the corona, and hence more works focused on comparing models with observations. \cite{Vial2016} compared their 1D isothermal, isobaric slab model with IRIS data. They were able to reproduce the unreversed, low-intensity Mg~II~h and k profiles with a certain combination of parameters (see their Table.~1). However, their model was unable to reproduce the low values of the k-to-h ratio. \cite{Heinzel2015} introduced a simple condensation--corona--transition region (CCTR) in their 1D model and found that this made an improvement in how well the model compares to the observed Mg~II and H$\alpha$ lines versus the simpler isothermal-isobaric models (1D or 2D). Furthermore, they discussed how the dynamics of the model also play an important role in recreating the more complex profile shapes.
 
   \cite{Jenkins2023} were one of the first to use more complex atmospheric models of prominences. They also set the foundations for the development of \texttt{Promweaver}, a framework used in this work and described in more detail in Sect.~\ref{sec:Num_Met}. \cite{Jenkins2023} used their 3D magnetohydrodynamic (MHD) simulation of a self-consistently formed prominence in a flux rope \citep{Jenkins2022}. They discussed the method used to create their synthetic spectra in detail and provided an in-depth comparison of the resulting filament and prominence synthetic profiles in different lines that are often used to observe cool plasma. \cite{Pietrow2024} and \cite{Snow2025} continued work synthesising spectra of complex atmospheres, both using \texttt{Promweaver}.

   We extended this work by performing a detailed analysis of how the synthetic spectra form from a complex 2.5D MHD simulation of highly dynamic, thread-like condensations resulting from a stochastic heating \citep{Jercic2024} and compare them to observational data. We give an overview of the simulation and a description of the method used to form the synthetic spectra in Sect.~\ref{sec:Num_Met}. In Sect.~\ref{sec:Observations} we describe the observations we chose to compare to this simulation, and then in Sect.~\ref{sec:results} we analyse how and why the atmosphere in the simulation results in specific synthetic spectra and how these compare to IRIS observations. In Sect.~\ref{sec:discors} we discuss our results and, lastly, in Sect.~\ref{sec:conclusion} we make conclusions based on the presented results.
  
\section{Numerical methods}
\label{sec:Num_Met}

\begin{figure}
    \centering
    \includegraphics[width=\linewidth]{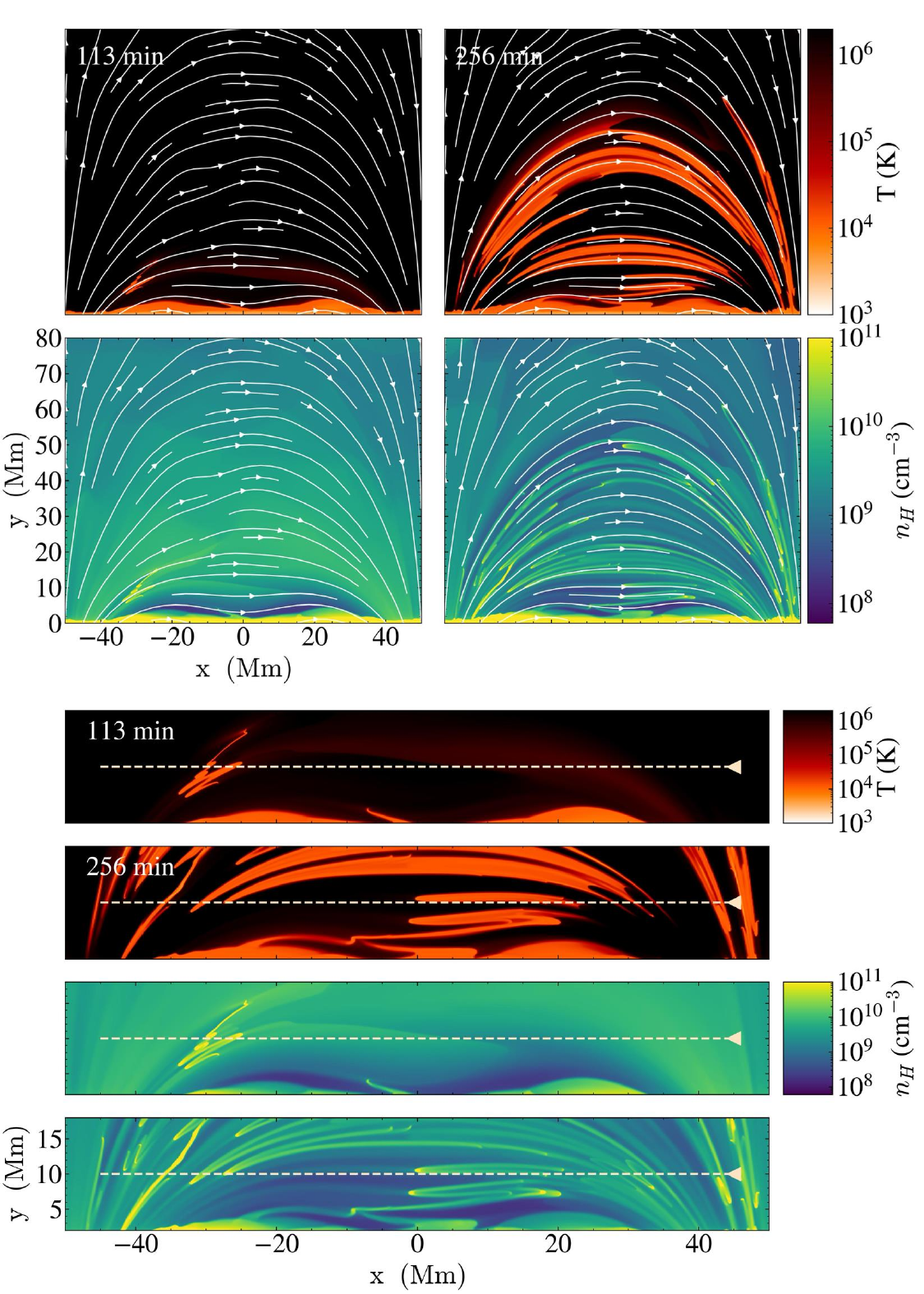}
    \caption{Temperature and density in the simulation, with bright colours showing colder and denser structures. At $t=113$\,min we show a moment early on in the evolution, while at $t=256$\,min we show a moment far into the simulation, when the domain was largely filled with condensation. The bottom four panels focus on the narrow region in which we took the cut along the dashed line representing the LOS and the arrow pointing in the direction of observation.}
    \label{fig:cond_sim}
\end{figure}
Our goal is to compare synthetic spectra created from a complex MHD simulation with real observations. We analysed a simulation described in \cite{Jercic2024}, whose authors presented two simulations that are a result of different types of localised heating \citep[a key ingredient to a thermal runaway process; for more details, see ][]{Antiochos_Klimchuk1991, Karpen2001, Xia2011}. For this study we focused on the simulation resulting from stochastic heating (pulses that are random in time and space located at the foot points of the magnetic topology supporting the dense plasma). For clarity, we repeat here the heating function used in \cite{Jercic2023}:
\begin{align}
    \label{eq:Hi}
        & H_i(t,x,y) =\nonumber \\
        &\begin{cases}
        E_1\sin\bigg(\frac{\pi(t-t_i)}{\delta t_i}\bigg)\exp\bigg(\frac{-|x-x_i|}{x_h}\bigg)\exp\bigg(\frac{-|y-y_i|}{y_h}\bigg), \hspace{0.3cm} t_i < t < t_i + \tau_i \\
        0 \hfill \text{otherwise,}
        \end{cases}
\end{align}
where $\delta t_i$ is the pulse duration and $x_h$ and $y_h$ are the heating length scales for x and y, respectively; both are 2\,Mm. The amplitude of this heating is defined as $E_1 = A(1 + \frac{\tau_i}{\delta\tau_i})$, where $A=0.2$\,erg\,cm$^{-3}$\,s$^{-1}$ and $\tau_i$ determines the interpulse duration. The total heating due to nanoflares is represented by
\begin{equation}
\label{eq:Htot}
    H=\sum_{i=1}^{n} H_i(t,s)\,.
\end{equation} 

We show two snapshots at different moments of the evolution in Fig.~\ref{fig:cond_sim}. To remain consistent with \cite{Jercic2024}, we mark the vertical direction on the $y$-axis. The simulation was performed with the state-of-the-art MHD code \texttt{MPI-AMRVAC} \footnote{\url{www.amrvac.org}} \citep{Xia2018, Keppens2020, Keppens2023}, where non-adiabatic effects such as radiative cooling, thermal conduction, and heating were considered. The heating is composed of the mentioned localised and background heating. The background heating, $H_{bg,}$ is defined as a simple exponential function, $H_0\exp(-\frac{y}{\lambda_0})$, with $H_0 = 10^{-4}$\,erg\,cm$^{-3}$\,s$^{-1}$, and $\lambda_0 = 50$\,Mm. For more details about the simulation and its setup, we refer the reader to \cite{Jercic2024}. About 100\,min after the localised heating was introduced, the first condensation starts to form (Fig.~\ref{fig:cond_sim}). As a result of the particular localised heating used in that simulation, the condensation does not form as a single monolithic blob of plasma, but rather as many highly dynamic threads. Due to highly varying heating and the particular arched configuration of the magnetic field, the threads in general drain quickly, and are just as quickly replenished by new condensations (although there is one obvious case of a thread persisting for roughly an hour). In \cite{Jercic2024}, the authors showed similarities with what is usually observed (the growth rate observed in prominences, dynamics, transverse oscillations of the threads, topological similarities). We took this comparison a step further and focused on creating and comparing the synthetic spectra from this simulation to observations of IRIS.

In order to create synthetic spectra from \texttt{AMRVAC} simulation, we used \texttt{Promweaver}\footnote{\href{https://doi.org/10.5281/zenodo.6546677}{Zenodo}}. \texttt{Promweaver} is a 1.5D code for solving radiative-transfer equations in statistical equilibrium with non-LTE conditions. It is a package built on top of \texttt{Lightweaver} \citep{Osborne2021}, a framework (rather than a 'code' due to its flexibility and ease of experimentation) made to simulate more general atmospheric setups, while \texttt{Promweaver} is specifically written to consider prominence (and filament) conditions. The particular 1.5D geometry means that \texttt{Promweaver} considers the properties of the atmosphere in only one coordinate and keeps them invariant in the other two directions \citep{Jenkins2023}. We describe only the most relevant facts about creating the synthetic spectra, but for further discussion and a more detailed overview, we refer the reader to \cite{Jenkins2023}. \texttt{Promweaver} simulates incident radiation based on the semi-empirical model of \cite{Fontenla1993}, more specifically their C model (often abbreviated as FALC). Any other pre-synthesised lower atmosphere could also be used with its own specific benefits and drawbacks. This represents the chromospheric model in the simulation, implemented as part of the boundary conditions of \texttt{Promweaver}. This was done in such a way as to prevent 'radiation trapping' \citep{Paletou1993}, which results from the constraints of using a 1.5D code. The limb darkening was also implemented as part of the boundary conditions, with the incident radiation being pre-computed for different incoming rays and gradually disappearing for angles much larger than the vertical  \citep[for more details on the implementation, see Sect. 2. in][]{Jenkins2023}.

To compute the synthetic spectra of a prominence view (the structure seen above the solar limb), we used the \texttt{StratifiedPromModel} of \texttt{Promweaver} and fed it with \texttt{AMRVAC} parameter values. We took a cut from the \texttt{AMRVAC} simulation at $y = 10$\,Mm and for $x \in [-45,45]$\,Mm. This is our line of sight (LOS), with the observer located on the right-hand side (positive part of the $x-$axis) of the simulation domain. We extracted the values of temperature, LOS velocities (in this case along $x$), and pressure. We chose this specific height in order to capture the first threads that formed and due to the sparsity of the tables of \cite{Heinzel2015}. As \texttt{AMRVAC} considers fully ionised plasma with a fixed 10:1 hydrogen-to-helium ratio, we calculated the number density, $n_e$, considering an ionisation degree, $i$:
\begin{equation}
    n_e = \frac{p}{(1+\frac{1.1}{i})k_BT}\,.
\end{equation} 
We inferred this ionisation degree using the tables of \cite{Heinzel2015} and considering the temperature and pressure values of \texttt{AMRVAC}. The initial population levels were determined by \texttt{Promweaver}  according to LTE. We further calculated the turbulent velocity according to Eqs.~(13)-(16) of \cite{Heinzel2001}. We considered Mg, H, and Ca as atoms in non-LTE, where a ten-level or higher continuum was used to model Mg~II, and a five-level or higher continuum was used for H and Ca~II (default atoms found in the \texttt{Promweaver} package). The line formation of these elements is treated assuming PRD \citep{Heasley_Kneer1976, Heinzel1987, Auer1994, Paletou1993}. For the results presented here, we used \texttt{Lightweaver v0.13.0} and \texttt{Promweaver v0.4.0}.

\section{Observations}
\label{sec:Observations}
\begin{figure*}
    \centering
    \includegraphics[width=0.9\textwidth]{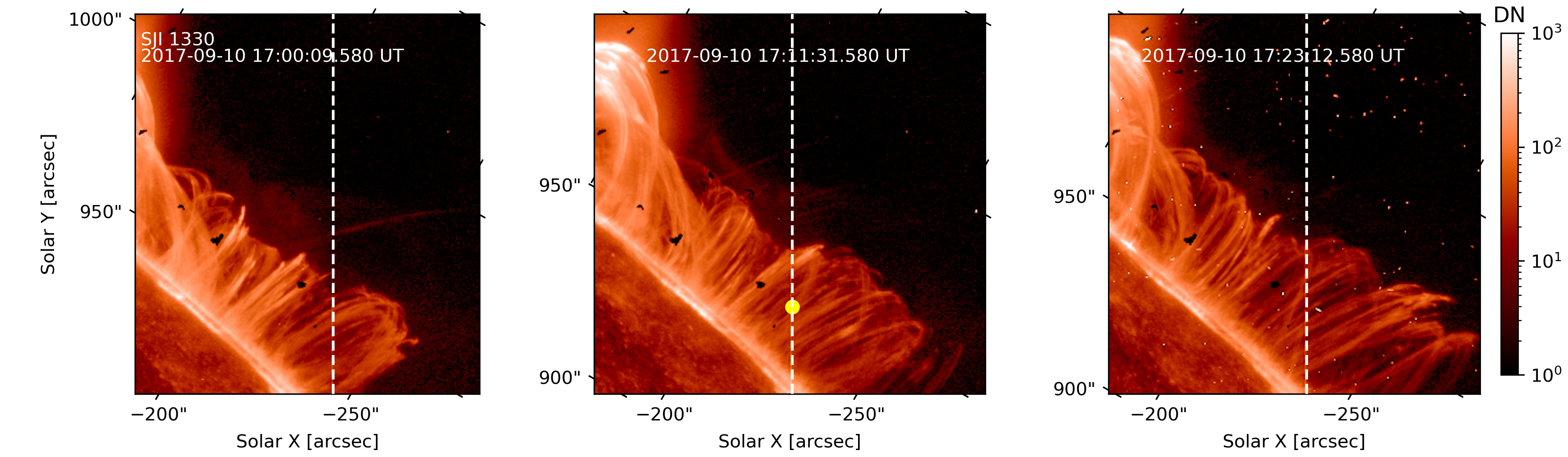}
    \caption{Post-flare coronal rain in September 2017. The dashed white line represents the position of the slit. The middle panel shows the location (yellow point) on the zeroth slit in the raster for the quartile analysis (see Sect.~\ref{subsec:quartile_ana}).}
    \label{fig:iris_Sep2017}
\end{figure*}

\begin{figure*}
    \centering
    \includegraphics[width=0.9\textwidth]{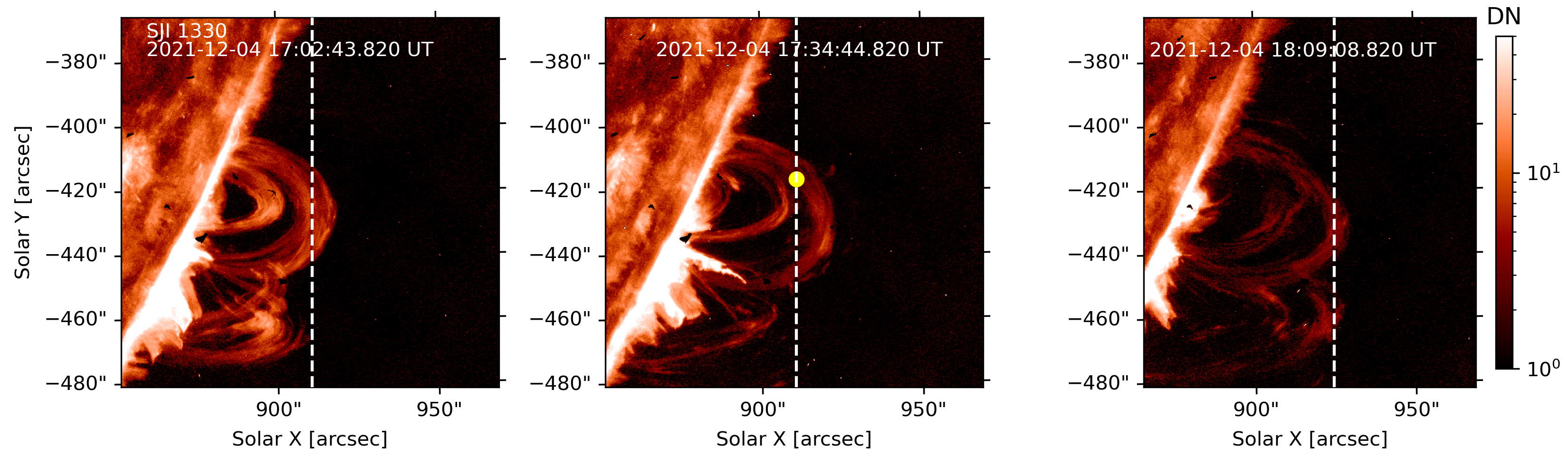}
    \caption{Same as Fig.~\ref{fig:iris_Sep2017} but for December 2021 data, again showing post-flare coronal rain. The middle panel shows the location (yellow point) on the zeroth slit in the raster for the quartile analysis (see Sect.~\ref{subsec:quartile_ana}).}
    \label{fig:iris_Dec2021}
\end{figure*}

\begin{figure*}
    \centering
    \includegraphics[width=0.9\textwidth]{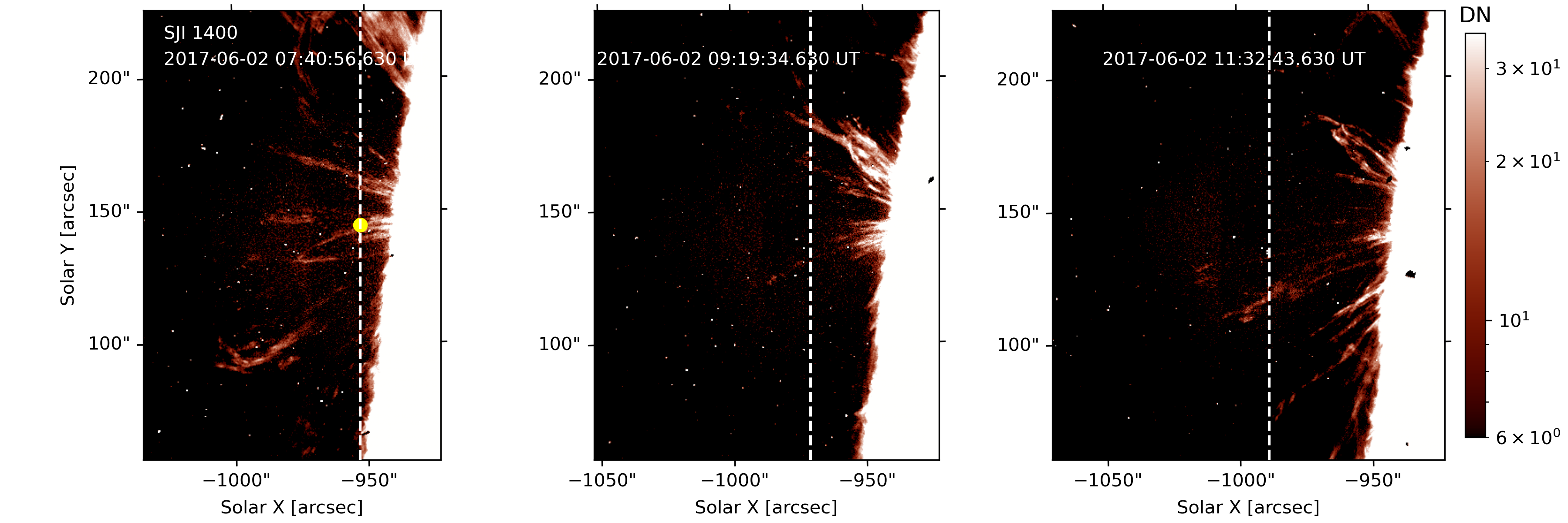}
    \caption{Same as Fig.~\ref{fig:iris_Sep2017} but for June 2017 data, showing quiescent, active-region coronal rain. The first panel shows the location (yellow point) on the 52nd slit in the raster for the quartile analysis (see Sect.~\ref{subsec:quartile_ana}).}
    \label{fig:iris_Jun2017}
\end{figure*}

Our choice of observations for comparison was guided by the simulation described in Sect.~\ref{sec:Num_Met}. The simulation was set up to recreate a thread-like structure of a prominence. However, due to the transient nature of the threads that form and the loop-like shape of the magnetic field, in its appearance the simulation more closely resembles observations of coronal rain. Searching for observations that first and foremost have topological similarities to our simulation, we settled for two observations made by IRIS, one on 10 September 2017 and the other on 4 December 2021, which in the rest of the paper we refer to as the September 2017 and December 2021 observations (Table~\ref{table:obs_details}). Both of these observations are flare-driven coronal rain. In order to avoid comparing only to such particular coronal rain events, we also looked into observation of quiescent coronal rain \citep{Antolin_Froment2022} above an active region. The chosen event happened on June 2 2017 and was first analysed by \cite{Sahin2022} and \cite{Sahin2023} with the IRIS/SJI pass bands. 
Details of the observations are given in Table~\ref{table:obs_details}. For the analysis we only used the raster data, but for context we show the accompanying slit-jaw images (SJIs) in Figs.~\ref{fig:iris_Sep2017},~\ref{fig:iris_Dec2021}, and~\ref{fig:iris_Jun2017}. For September 2017 and December 2021 the SJIs are, in both cases, 1330\,\AA~with a FOV of 119"x119", while for June 2017, we show 1400\,\AA~with a FOV of 167"x175". The amount of noise is very high for SJI of June 2017, so we only plot the SJI above 10\,DN to alleviate the influence of this noise. For the other two observations there is less noise, and we plot the SJI above 1\,DN. Notice that the colour-bar upper limits are different for each observation, as each has a different intensity. However, the panels for each image share the same colour bar, and this reflects the change in intensity with the evolution of the rain in each event. This is particularly noticeable in Fig.~\ref{fig:iris_Dec2021}, in which the rain slowly fades away with time. The raster data were retrieved as level 2 fits files, where they are given in counts or data number units (DNs)\footnote{\url{https://iris.lmsal.com/itn26/itn26.pdf}}. 

On 10 September 2017, the NOAA active region 12673 produced four smaller C-class flares (from about 3:00 until 15:00 UT) until 16:06 UT, when an X-class flare reached its peak \citep[according to the data of the XRT flare catalogue;][]{Watanabe2012}. About half an hour later coronal rain was observed by IRIS. The IRIS observation started at 12:59 and ended at 19:23\,UT, and we show a few relevant snapshots of the evolution of the post-flare coronal rain in Fig.~\ref{fig:iris_Sep2017}. 

According to the XRT catalogue, only C- and B-class flares were observed on 4 December 2021 from 3:30 until 23:59 UT \citep[according to the XRT flare catalogue;][]{Watanabe2012}. The observation of December 2021 started at 17:02 and ended at 18:09\,UT. It also represents a post-flare coronal rain. However, there were no significant strong flares that occurred in the close period prior to the rain. The IRIS observations do not capture the particular moment of the formation of this rain (as they do for the September 2017 case). We show the evolution of this coronal rain in Fig.~\ref{fig:iris_Dec2021}. From these snapshots it is obvious that the coronal rain slowly extends in height, but it also fades from the 1330\,\AA~channel in the one hour of evolution captured by IRIS.

\begin{table*}
      \begin{threeparttable}[b]
    \captionsetup{justification=centering}
    \caption{Location and timing of September 2017 and December 2021 IRIS observations.} 
    \label{table:obs_details}      
    \centering                          
    \begin{tabular}{c|c|c|c|c|c|c|c}        
    \hline\hline                
    Dataset & Coord. & Start - End & Max & Raster & \multicolumn{2}{c|}{Cadence [s]} & CDELT1\&2 ["]\\  
     & (xcen,ycen) & [UT] & FOV & Steps & Raster & SJI & (SJI) \\
    \hline
    \hline
     \href{https://www.lmsal.com/hek/hcr?cmd=view-event&event-id=ivo%3A%2F%2Fsot.lmsal.com%2FVOEvent%23VOEvent_IRIS_20170910_125947_3660109533_2017-09-10T12%3A59%3A472017-09-10T12%3A59%3A47.xml}{June 2017} & (-970",143") & 2017-06-02 & 230"$\times$175" & 64$\times$1" & 16.2 & 43 & 0.33\\
     & & 07:28:00-12:55:43 & & & & & \\
    \hline   
     \href{https://www.lmsal.com/hek/hcr?cmd=view-event&event-id=ivo%3A%2F%2Fsot.lmsal.com%2FVOEvent%23VOEvent_IRIS_20170910_125947_3660109533_2017-09-10T12%3A59%3A472017-09-10T12%3A59%3A47.xml}{September 2017} & (981", -209") & 2017-09-10 & 133$\times$119 & 8$\times$2" & 9.3 & 9 & 0.33\\  
     & & 12:59:47-19:23:38 & & & & & \\   
    \hline
     \href{https://www.lmsal.com/hek/hcr?cmd=view-event&event-id=ivo%3A%2F%2Fsot.lmsal.com%2FVOEvent%23VOEvent_IRIS_20211204_170243_3660259533_2021-12-04T17%3A02%3A432021-12-04T17%3A02%3A43.xml}{December 2021} & (918",-422") & 2021-12-04 & 133"$\times$119" & 8$\times$2" & 9.6 & 77 & 0.16 \\ 
     & & 17:02:43-18:09:18 & & & & & \\
    \hline
    \hline
    \end{tabular}
      \end{threeparttable}
    \end{table*}

\section{Results}
\label{sec:results}

We focused chiefly on the Mg~II~k line, as it is representative for both resonance lines of Mg~II. We discuss the results of forward modelling using \texttt{Promweaver} and go into the details of the synthetic spectra and how well they compare to observations.

\subsection{Formation of the condensation through the eyes of Promweaver (and the four panel plots)}
\label{sec:4pp}

\begin{figure}
    \centering
    \includegraphics[width=\linewidth]{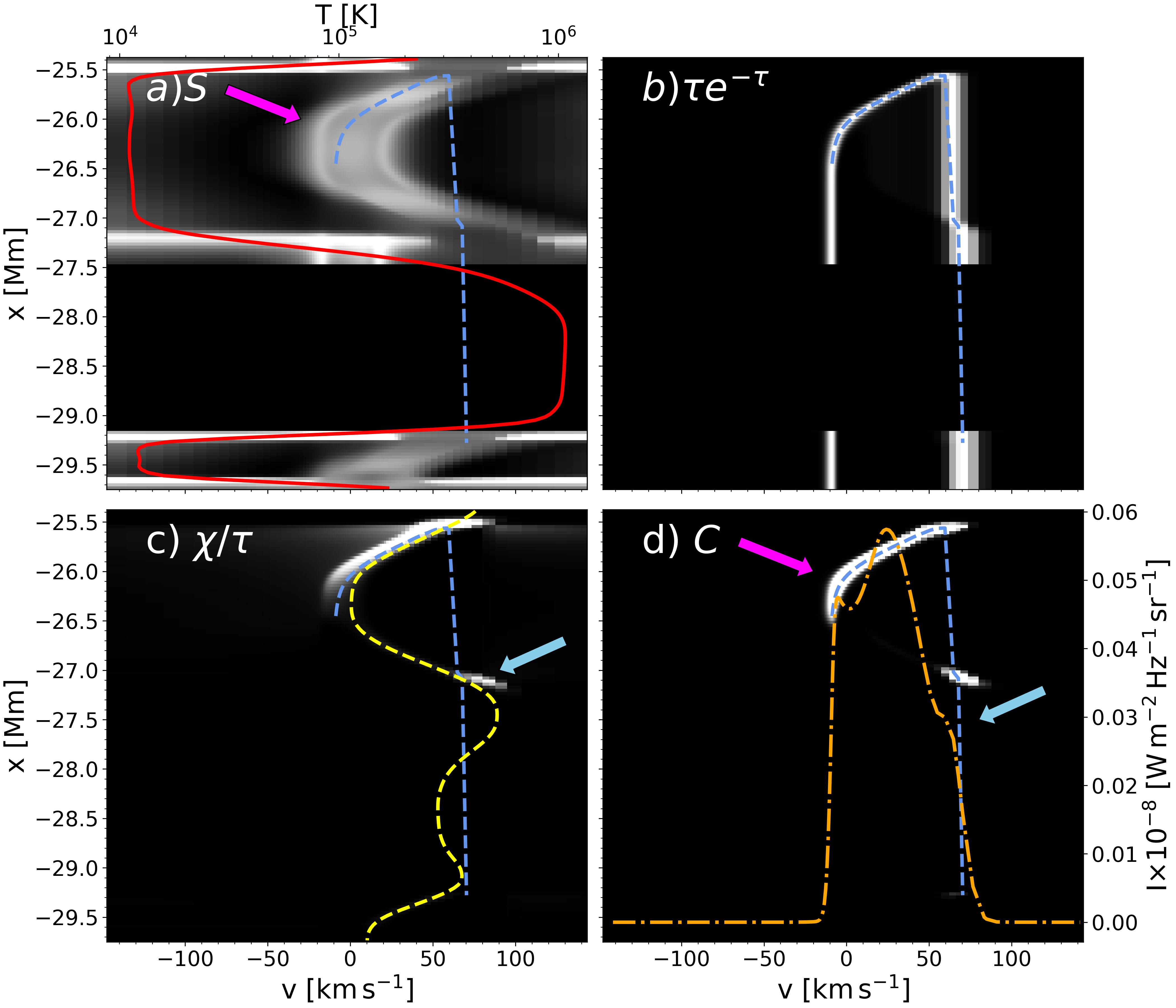}
    \caption{Moment just after the condensation forms (corresponding to the left panels in Fig.~\ref{fig:cond_sim}) at $t=113$\,min for which we show d) the contribution function, $C,$ and its factors; a) the source function, $S$; b) $\tau_{\nu}e^{-\tau_{\nu}}$ ; and c) $\chi_{\nu}/\tau_{\nu}$. The red profile in the upper left panel is the temperature along the LOS, the dashed blue line in all panels is tracing $\tau=1$ (with the observer being from the positive side of the $x$-axis), the dashed yellow line in the lower left panel is the LOS velocity ($x$ component of the velocity in \texttt{AMRVAC}, converted to the observer's reference frame), and the dash-dotted orange line in panel d) represents the spectral line with the corresponding intensity axis on the right. The arrows mark particular contributions and how they are seen on the spectral line (shown in d). The regions with $T>260\,$kK are masked in black and are excluded from the analysis.}
    \label{fig:4pp1}
\end{figure}

\begin{figure}
    \centering
    \includegraphics[width=\linewidth]{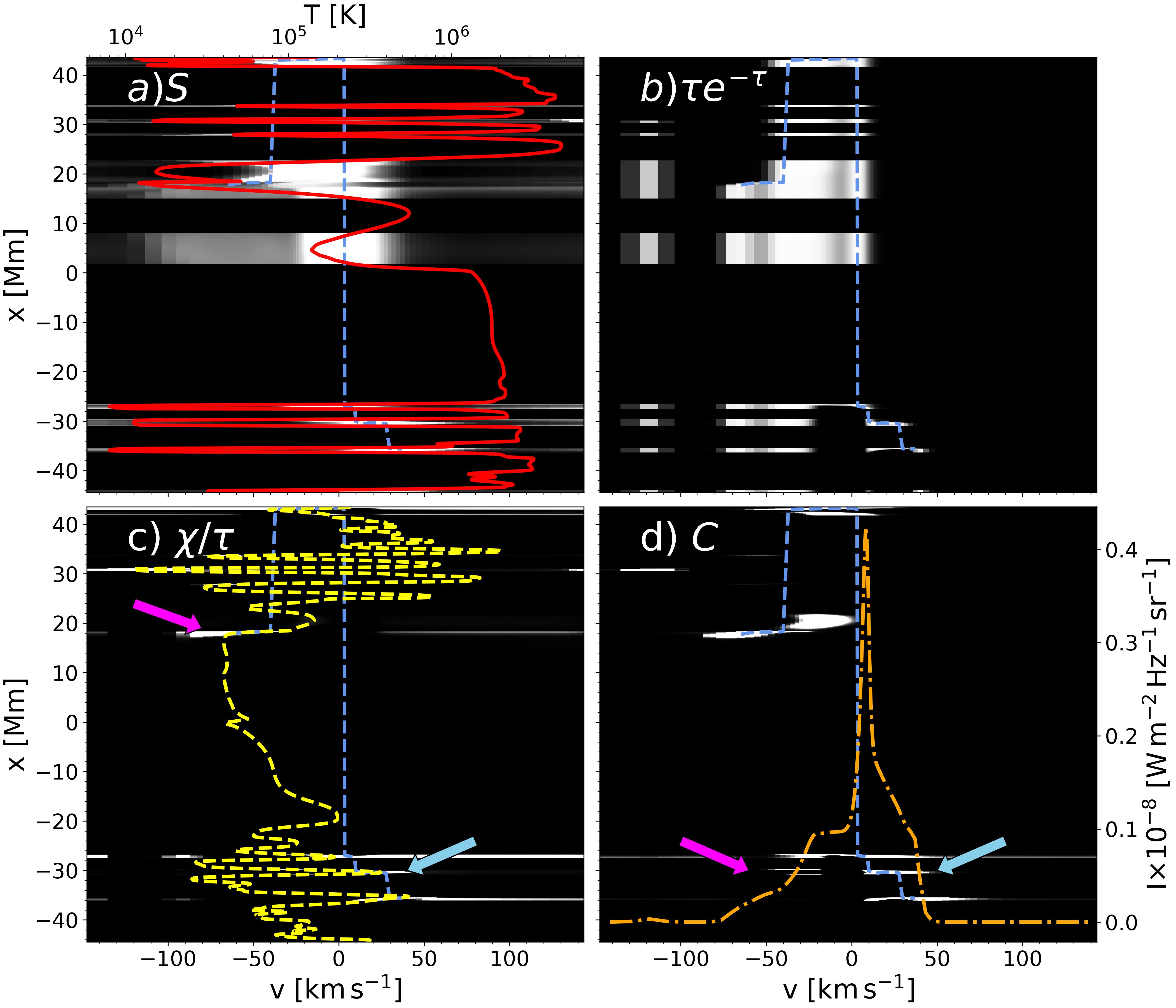}
    \caption{Same as Fig.~\ref{fig:4pp1} but later on in the evolution when many different threads along the LOS have formed (corresponding to the right panels in Fig.~\ref{fig:cond_sim} and bottom panel in Fig.~\ref{fig:q123_locations}).}
    \label{fig:4pp2}
\end{figure}

In order to understand in greater detail how the spectral lines form in the simulation, we started by applying the method of \cite{Carlsson_Stein1997}. We analysed the intensity integrand, which is better known as the contribution function $C$: 
\begin{equation}
    I_{\nu} = \int_{x_0}^{x_1} S_{\nu} \tau_{\nu} e^{-\tau_{\nu}} \frac{\chi_{\nu}}{\tau_{\nu}}dx = \int_{x_0}^{x_1} C(x) dx\,.
    \label{eq:cont_func}
\end{equation}
$S_{\nu}$ is the frequency dependent source function and $\tau_{\nu}$ is the optical depth (dependent on the wavelength and $x$-coordinate), where $e^{-\tau_{\nu}}$ is the attenuation factor. $\chi_{\nu}$ is the opacity, representing the density of the emitting particles. Equation~\ref{eq:cont_func} should also contain $\mu$, the viewing angle, which is in our case equal to one, so we omitted it. In Figs.~\ref{fig:4pp1} and~\ref{fig:4pp2} we show these terms. Panel a) shows the source function, $S_{\nu}$; b) shows the attenuation factor multiplied by $\tau_{\nu}$; and c) shows the ratio of the opacity and optical depth. Lastly, panel d) represents the multiplication of all these terms, that is, the resulting contribution function, $C$. These are the now-standardised four-panel plots, where we flipped the Doppler axis (blueshift meaning negative and redshift positive) to facilitate comparison to observations. In addition to the parameters shown in grey scale, additional lines have been added to help with the interpretation. The red profile in panel a) is the profile of the temperature along the LOS ($y=10$\,Mm). Panel c) has a dashed yellow profile, representing the $x$ component of the velocity profile from AMRVAC (converted to the observer's reference frame where negative means towards the observer and positive away from it, which is opposite to what it would be in the simulation). Panel d) has an dash-dotted orange profile, which represents the shape of the spectral line we analysed (with the corresponding intensity axis plotted on the right-hand side). All the panels also have a dashed blue line that traces the location of $\tau=1$. The lack of the line corresponds to the lack of $\tau$ reaching the value of one. We note that the observer is located on the positive side of the $x$-coordinate (i.e. from the top in these figures).

In order to reduce the computational load, we set a threshold temperature of $T>260$\,kK, above which \texttt{Promweaver} ignores the domain. The threshold value of 260\,kK was chosen arbitrarily. It is high enough for \texttt{Promweaver} to capture all the relevant (optically thick and thick to thin transitions such as the CCTR) regions and low enough to save on computation time. Hence, in all the panels here, we only plot the extent of the domain occupied by the cold threads, which can significantly differ at different moments (e.g. 4\,Mm in Fig.~\ref{fig:4pp1} vs. $>$\,80\,Mm in Fig.~\ref{fig:4pp2}). Because \texttt{Promweaver} ignores parts of the domain with $T>260$\,kK---yet with these locations sometimes occurring between adjacent condensations---we masked regions with those temperatures in black (blending with the background colour of the image); for example, between -27.5 and -29.1\,Mm in Fig.~\ref{fig:4pp1}, or the bigger part of the region between 16 and -26\,Mm in Fig.~\ref{fig:4pp2}. These regions are excluded from the analysis.

The conditions we analysed are non-linear and far from LTE (i.e. the source function is far from the Planck function); hence, the source function and other parameters shown in Fig.~\ref{fig:4pp1} and~\ref{fig:4pp2} exhibit an intricate and complex behaviour with LOS and wavelength. Figure~\ref{fig:4pp1} shows the state of the condensation early on in the evolution, when there is only one thread ($t=113$\,min, upper panel of Fig.~\ref{fig:cond_sim}). Hence, it is relatively straightforward to determine the source of the spectral line. The $y-$axis in Fig.~\ref{fig:4pp1} extends for only about 4 Mm, a narrow area encompassing the full source region of the spectral line. The location of $\tau=1$ indicates that the line is formed predominantly in the thread stretching from $-25.5$ up to $-27$~Mm. This region is narrow enough for us to analyse the change of the source function in this thread (marked by the magenta arrow). Following the blue line in panel a) of Fig.~\ref{fig:4pp1}, we see that in this region the source function varies (going from left to right) from slightly brighter to darker, to brighter again. This effect is reflected in the spectral line (dash-dotted orange line in panel d)), showing a small peak, then a decrease, and then a peak again (going from about -15 to 30\,km\,s$^{-1}$ and marked with a magenta arrow). The full width of the line is further extended by a condensation edge brightening up at about 70\,km\,s$^{-1}$ in the $\chi_{\nu}/\tau_{\nu}$ parameter (marked with the light blue arrow in panel c)). We see a corresponding slight bump in the spectral line (dash-dotted orange line) at $x\approx$70\,km\,s$^{-1}$ (marked with the corresponding light blue arrow in panel d)). The source function shows another interesting phenomenon. Despite showing increased values in the thread itself, it is comparatively much stronger for the CCTR (corresponding to regions with temperatures of about 20--30\,kK). However, these regions in the line wings are optically and spatially thin. In comparison with the emission coming from the core of the thread,  these regions only weakly contribute to the emergent intensity.

In Fig.~\ref{fig:4pp2}, we consider the second time period ($t=256$\,min) when there is much condensation along the LOS, resulting in a more complex source region. However, starting from panel b), we can determine which threads contribute to the spectral line. We see that $\tau=1$ only occurs in certain threads, and this results in the $\tau=1$ line (blue dashed) having a particular step-like shape rather than a smooth profile \citep[cf.][]{Carlsson_Stein1997}. Panel c) shows the ratio of opacity, $\chi_{\nu}$, and optical depth, $\tau_{\nu}$. As $\chi_{\nu}$ represents a measure corresponding to density and, similarly, $\tau_{\nu}$, both parameters are affected by the presence of a velocity gradient and create an asymmetry in the spectral line \citep{Carlsson_Stein1997}. We also see this in Fig.~\ref{fig:4pp1}, but with so many velocity gradients here it has a much stronger influence on the spectral line. Velocity gradients are related to locations of increased density, meaning large opacity in front of the gradient and small optical depth behind it; hence, $\chi_{\nu}/\tau_{\nu}$ will peak and contribute more significantly to the intensity we observe. The thread at about $x=18$\,Mm, where the $\tau=1$ line starts, has an increase in the $\chi_{\nu}/\tau_{\nu}$ value, which can be related to the velocity gradient at that particular location (of about -70~km\,s$^{-1}$). This particular contribution adds to the increased width of the spectral line, increasing the intensity in the far blue wing (magenta arrows in panels c) and d)). The thread is blueshifted because it is moving towards the observer and falling back towards the chromospheric layers. Another blueshifted thread contributing to the extension of the line's width is the one at $x=42$~Mm (closest to the observer), for which we also see an increase in the $\chi_{\nu}/\tau_{\nu}$ parameter. Further on, we see $\tau=1$ forming in a thread at about $-27$~Mm. Here, we again see an increase in the  $\chi_{\nu}/\tau_{\nu}$ (related to a velocity gradient of a couple of km\,s$^{-1}$ and corresponding to the redshift of the main peak). However, we also see an increase in the source function for the short range of $-26$ to $-28$~Mm (corresponding to the region of low temperature,~10~000\,K. Since there is a contribution from the source function and the  $\chi_{\nu}/\tau_{\nu}$, we see a resulting strong peak in the Mg~II~k line. There are further contributions by threads at about $-30$ and $-36$~Mm. Both threads show an increase in both the $\chi_{\nu}/\tau_{\nu}$ (velocity gradients at about 30 and 40~km\,s$^{-1}$) and in the source function (marked by light blue arrows in panels c) and d)). Their brightness in the $\chi_{\nu}/\tau_{\nu}$ (due to the velocity gradient) is much less than for the one caused by the central peak, and, as such, these threads merely add to the extension of the line width in the far red wing, but there are no more pronounced peaks in the spectral line.

\subsection{Comparison of synthetic and observed spectra}
\label{subsec:quartile_ana}
\begin{figure}
    \centering
    \includegraphics[width=0.86\linewidth]{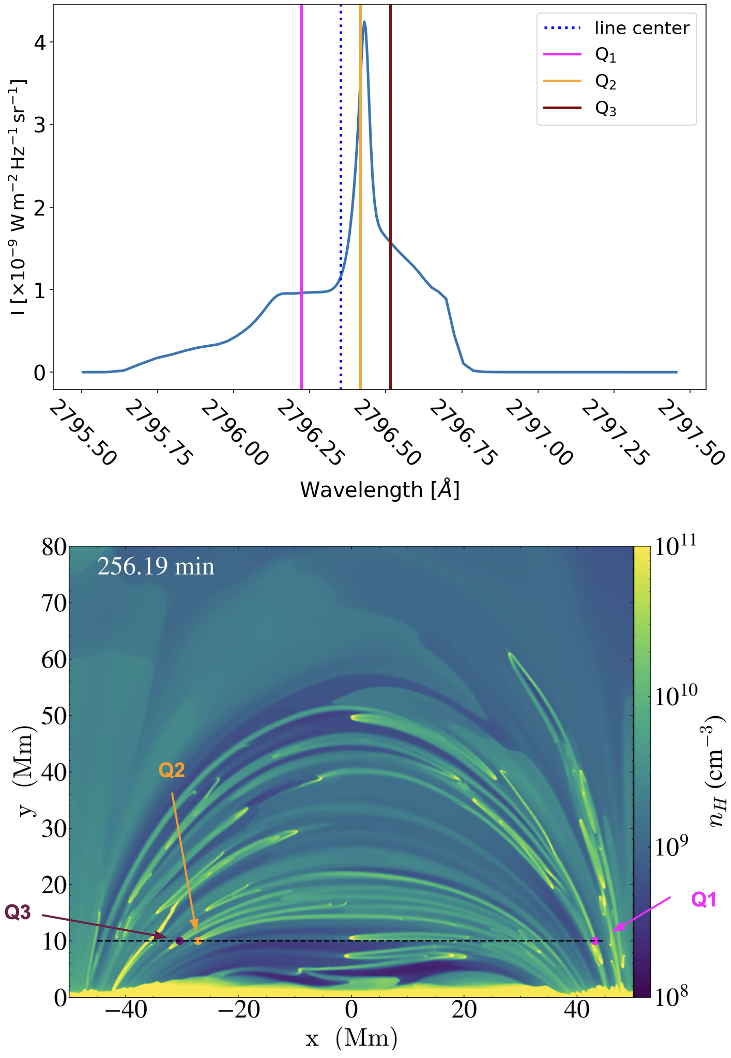}
    \caption{Moment in the simulation ($t=256$\,min) showing how different parts of our LOS contribute to the spectral line. The upper panel shows the spectral line in question, with different quartiles ($Q_1$, $Q_2$ and $Q_3$) marked with corresponding colours, together with the theoretical line centre (dotted blue line at 2796.35\,\AA). The lower panel shows the density structure in the simulation with the LOS (dashed black line) and the locations of $\tau=1$ for the corresponding quartile wavelengths.}
    \label{fig:q123_locations}
\end{figure}

\begin{figure}
    \centering
    \includegraphics[width=\linewidth]{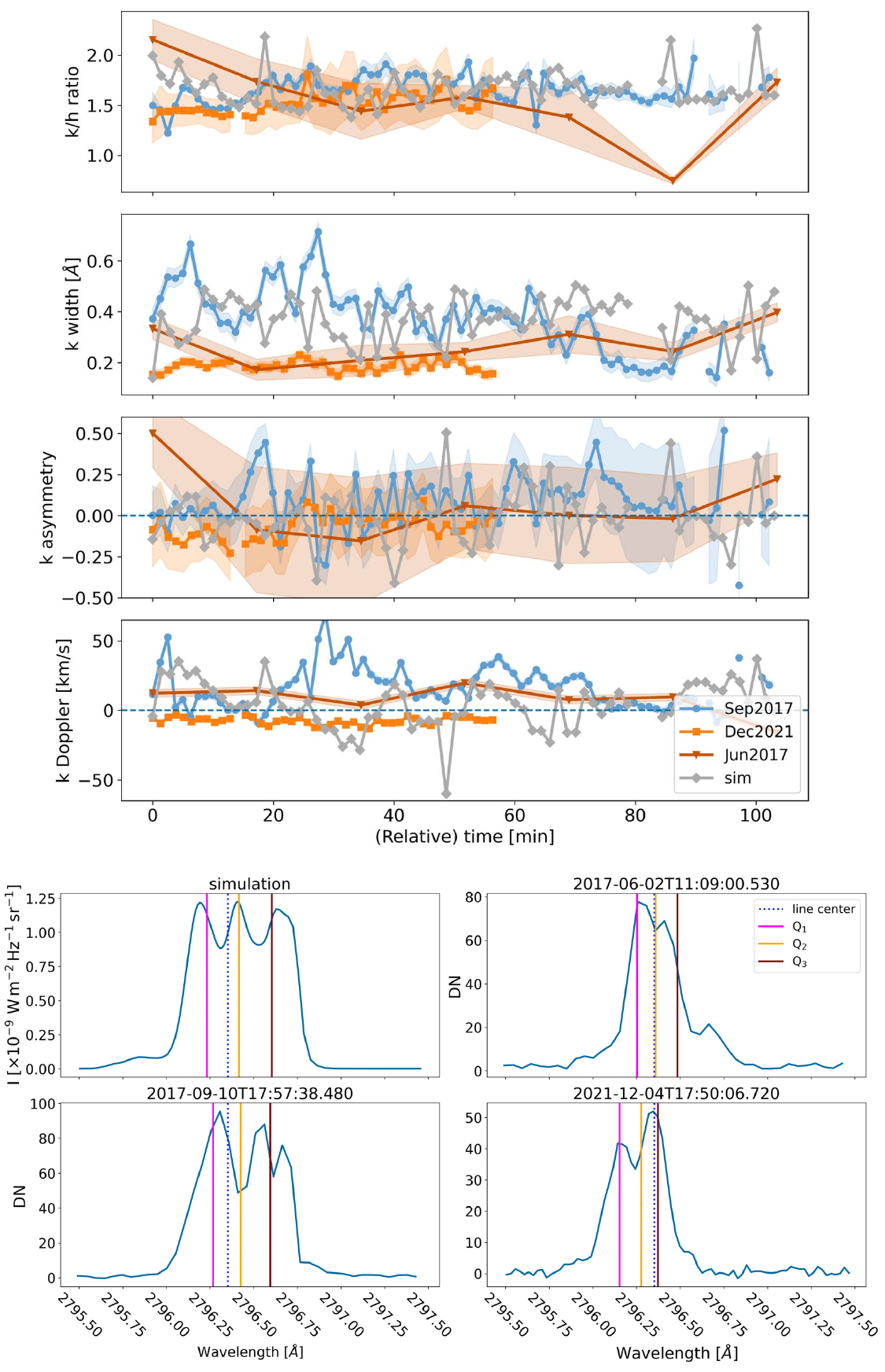}
    \caption{Upper four panels show the time evolution of observable parameters (k/h ratio, width, asymmetry, and Doppler shift) for the simulation and together with the derived 1$\sigma$ standard deviation (shaded region) for IRIS observations. The dashed lines in the two bottom panels mark zero values. December 2021 is shown from the start of the IRIS observation, June 2017 starts at 08:16\,UT, and September 2017 starts from the appearance of coronal rain at about 16:25\,UT. The four additional panels at the bottom show examples of spectra and their corresponding quartiles (Q$_1$, Q$_2$, Q$_3$), with the vertical dashed line showing the line centre (2796.35\,\AA).}
    \label{fig:quartile_analysis}
\end{figure}

In this section, we describe how observable parameters evolve in time for the case of simulation and observations. To achieve a quick and reliable summary of the spectral behaviour, we opted for the so-called quartile analysis. The Mg~II lines show complex behaviour (including single, double, and even more multiply peaked spectra; see examples in Figs.~\ref{fig:q123_locations} and~\ref{fig:quartile_analysis}). The particular shape of the spectral line seen in Fig.~\ref{fig:q123_locations} is an example where the dynamics of the threads with respect to the incident radiation has a strong influence \citep[see Fig.~5 in][]{Gunar2024}. We calculated the observables following the method of \cite{Kerr2015} \citep[also used in][]{Ruan2018, Peat2024_observations}. To analyse only the significant spectral lines and to be sure we were not analysing noise, we first checked for peaks in the range of 2\,\AA~around the centre of the Mg~II~k and h lines (2796.35 and 2803.53\,\AA, respectively). If there was no peak greater than 25\,DN in that range, we did not consider that specific snapshot.

The quartile analysis consists of searching for 25, 50, and 75\% quartiles of the cumulative distribution of intensity. We refer to the wavelengths that correspond to these values as $Q_1$, $Q_2$, and $Q_3$, respectively. As in \cite{Kerr2015}, we used these values to calculate
\begin{align}
    &\lambda_c = Q_2\,, &\text{line centroid position;} \label{eq:q1} \\
    &W = Q_3 - Q_1\,, &\text{width of the line;} \label{eq:q2} \\
    &S = \frac{(Q_3-Q_2)-(Q_2-Q_1)}{Q_3-Q_1}\,, &\text{asymmetry.} \label{eq:q3} \,
\end{align}
We used the position of the centroid, $\lambda_c$, to calculate the Doppler shift:
\begin{equation}
\label{eq:doppler}
    v_{LOS} = \bigg(\frac{\lambda_c}{\lambda_0} - 1\bigg)c\,.
\end{equation}
As we focused on the Mg~II~k line, $\lambda_0$ represents the theoretical position of the centre of the Mg~II~k line (in \texttt{Promweaver} defined as 2796.35\,\AA) and $c$ is the speed of light. An example of how the quartiles are located in the domain is shown in Fig.~\ref{fig:q123_locations}, corresponding to the same time moment as Fig.~\ref{fig:4pp2} of $t=256$\,min. In the upper panel we see the spectral line with its $Q_1$, $Q_2,$ and $Q_3$ wavelengths marked by vertical lines. In the lower panel we show the corresponding snapshot of the simulation (the density distribution) and the location where each of these parts of the spectra is formed ($\tau=1$ for each particular wavelength).

Figure~\ref{fig:quartile_analysis} shows the time evolution of these parameters for the simulation and in a similar way chosen for one pixel, at one slit position, for observations of June 2017, September 2017 and December 2021 (the particular pixel chosen for this analysis is shown in the accompanying SJIs in Figs.~\ref{fig:iris_Sep2017},~\ref{fig:iris_Dec2021}, and~\ref{fig:iris_Jun2017}.) 
The missing simulation values happen because \texttt{Promweaver} did not converge for the atmosphere at that time. The missing observation values are either due to the peak being smaller than 25\,DN or due to erroneous pixels (during times of high energetic particle flux). An advantage of these three observations is that each provides us with a different LOS with respect to the coronal rain motion. This influences the changes seen in Fig.~\ref{fig:quartile_analysis}. In the case of September 2017, our LOS is roughly in line with the coronal rain motion (though not exactly). This is similar to the simulation, where the LOS is perfectly parallel with the motion of the moving threads (the threads fall along $x,$ and our LOS is along the same coordinate). On the other hand, the observation of December 2021 has a LOS that is (roughly) perpendicular to the plane of the sky (POS) in which the coronal rain is falling. For June 2017 and the particular pixel we chose, from \cite{Sahin2023} we know that the mean spatial angle is $\sim0^{\circ}$ with respect to the solar radial direction \citep[see Fig. 2 in][]{Sahin2023}. Additionally, because of the expected shape of the magnetic field around a sunspot and the fact that we captured motion relatively close to the solar surface, we predominantly captured vertically falling threads. This provides us with yet another perspective on the motion of the coronal rain.

In Fig.~\ref{fig:quartile_analysis} the upper panel shows the k/h ratio. The parameter is calculated as the ratio of integrated intensities in the range of 2\,\AA~(from 2802.7 to 2804.7~\AA~ for Mg~II~h and from 2795.5 to 2797.5~\AA~for Mg~II~k). The value of the k/h ratio indicates if the source is optically thin or thick in these lines \citep{Kerr2015}. In all three cases presented, the k/h ratio is $<2$ overall, indicating that we are looking at predominantly optically thick structures. December 2021 shows a significantly smaller width (varying in a very narrow range at $\sim$0.2\,\AA), asymmetry (around 0), and Doppler shift (around 0~km~s$^{-1}$) than the observation of September 2017 and the simulation. September 2017 and the simulation show a much closer range of values. The width varies between 0.2 and 0.6\,\AA, while the asymmetry is broadly distributed within $\pm0.5$ with sharp changes in between these values. The Doppler shift varies in a wide range of $\pm50$\,km\,s$^{-1}$. For June 2017 we only have a point every 17\,min. Its k/h ratio shows the largest range of values going from 0.75 up to 2.15. The width shows smaller values in comparison to September 2017, remaining in the range of approximately 0.2 and 0.4\,\AA; however, still larger than those of December 2021 case. If we put aside the first point in the dataset, which corresponds to the largest k/h ratio value in the range (leaning more towards optically thin plasma), asymmetry is relatively small, varying between -0.15 and 0.22. The Doppler shift does not vary strongly and is in the range of -16 and 20\,km\,s$^{-1}$, with an average of 7\,km\,s$^{-1}$.

The lowest four panels in Fig.~\ref{fig:quartile_analysis} show examples of spectral lines in all three observations and the simulation together with the corresponding quartiles (marked by the full lines). The spectra can vary greatly throughout the simulation; they can be single peaked, double peaked, and even have more complex shapes, as demonstrated here. This particular example shows that the synthetic spectra successfully recreate even some of the more complex spectral lines we see in the observations.

\subsection{Correlations based on the simulation}
\label{sec:Correlations}
\begin{figure*}
    \centering
    \includegraphics[width=0.9\textwidth]{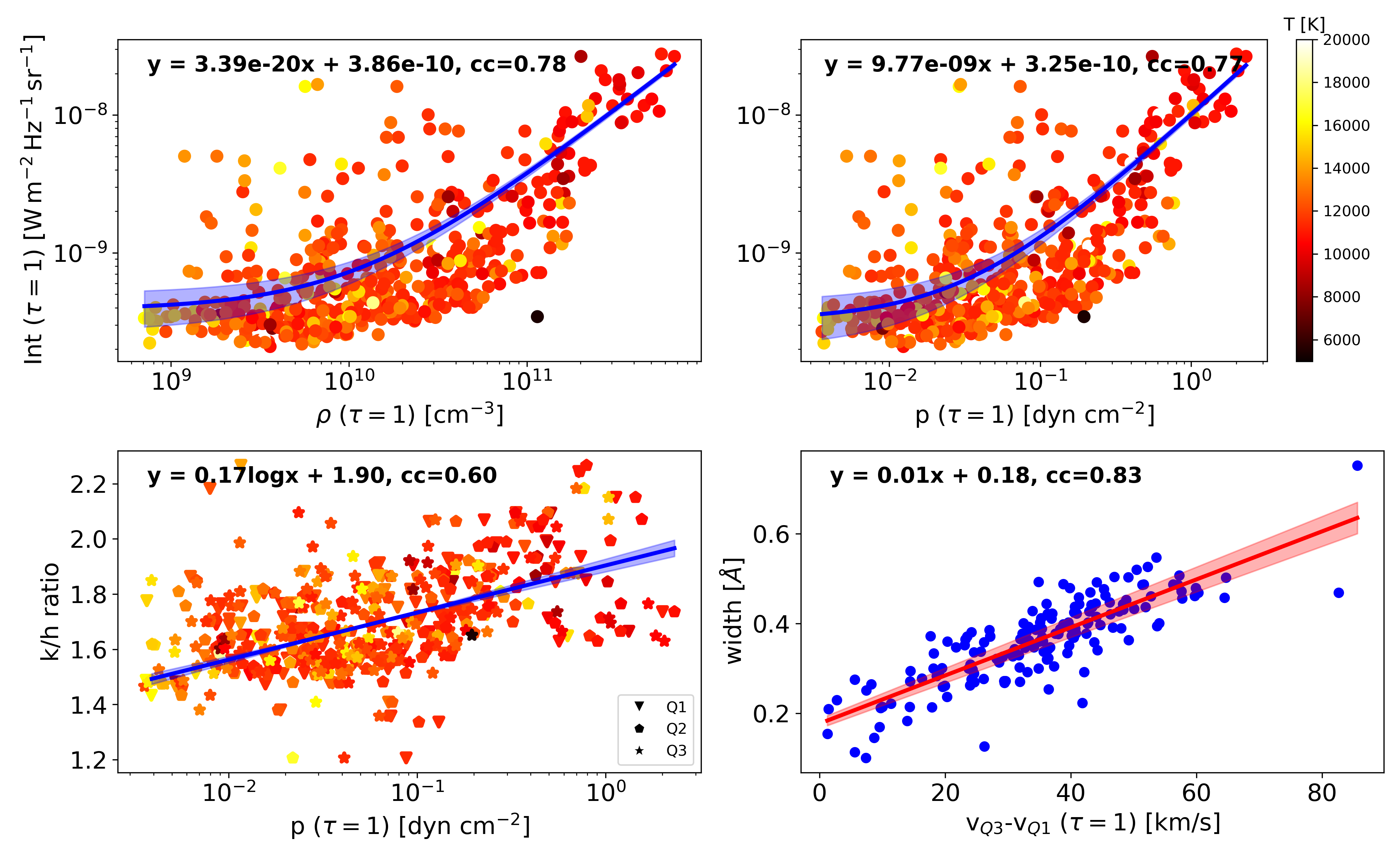}
    \caption{Correlations between observables as measured from the synthetic spectra and the atmospheric parameters of the simulation. The linear equation in the bottom left panel is for the k/h ratio and $logp$ only at $\tau=1$ of Q$_2$. For clarity we show the top two panels on a log--log scale.}
    \label{fig:correlations}
\end{figure*}
We want to explore the relation between the observables we measured from the synthetic spectra and the parameters of the simulation. Figure~\ref{fig:correlations} shows how the intensity, k/h ratio, and the width of the spectra correlate with density, pressure, and velocity, respectively. Density, pressure, and velocity were extracted from the simulation at the $x$-coordinate where $\tau=1$ of $Q_1$, $Q_2$, and $Q_3$ wavelengths. \cite{Jenkins2023} explored two different methods of calculating the formation height of a particular line. One method is locating the height where $\tau=1,$ and the other is calculating the average formation height weighted by the contribution function, $C,$ as given in Eq.~\ref{eq:cont_func}. They concluded that the two methods give similar results, though in the more optically thin cases they favour the average formation height approximation. As we dealt with predominantly optically thick lines, for the purposes of this study we kept using the location of $\tau=1$. 

Figure~\ref{fig:correlations} shows pressure and density with the corresponding intensity for every $Q_1$, $Q_2$, and $Q_3$. In the case of the k/h ratio, we plot the same k/h values for each set of pressures we extracted for $Q_1$, $Q_2$, and $Q_3$ (marked with different symbols in the bottom left panel). The width of the spectra is correlated with the difference of velocities calculated at $Q_3$ and $Q_1$. This relation is a result of different structures moving in opposite directions along the LOS, so the resulting spectra is a superposition of different Doppler shifts \citep{Gunar2008, Heinzel2015_letter, Peat2023, Schwartz2024}. Intensity and density have a Pearson correlation coefficient of 0.78 (Fig.~\ref{fig:correlations} shows the values on a log--log scale for clarity), k/h ratio, and $log_{10}p$ (only at Q$_2$) have a value of 0.60, and the width and velocity correlate with a Pearson correlation coefficient of 0.83. The shaded area around the fitted line represents one standard deviation. We note here that the upper two panels show a linear relation on a log--log scale for more clarity. For the same reason, the lower left panel only has the $x$-axis log-scaled. We further discuss these correlations in Sect.~\ref{sec:discors}.
\begin{figure}
    \centering
    \includegraphics[width=\linewidth]{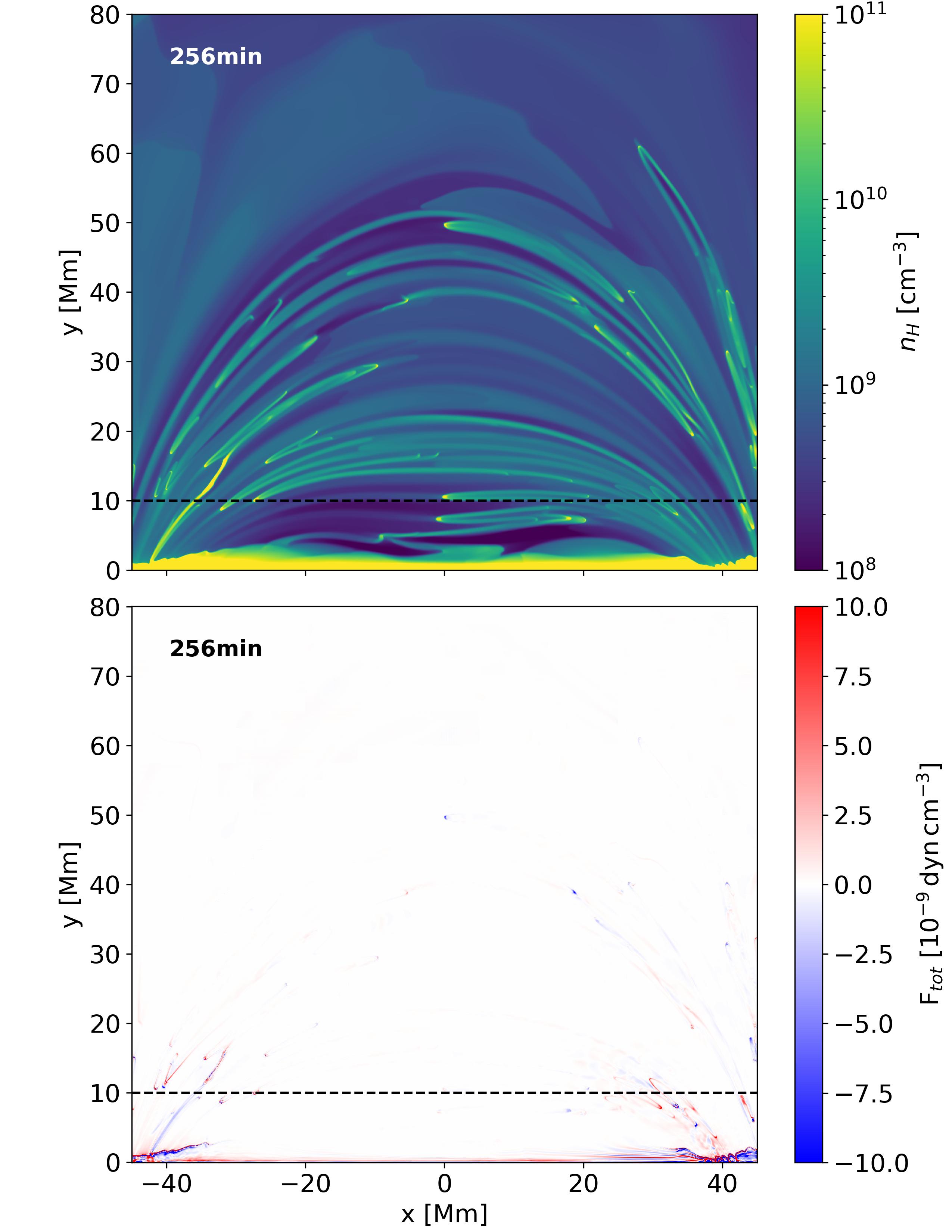}
    \caption{Upper panel shows the number density for one representative snapshot of the simulation ($t=256\,$min), with the dashed black line representing our LOS. The lower panel shows the same snapshot with the corresponding values of the sum of vertical components of the Lorentz force and the pressure gradient, with the gravity force being $F_{tot}$.}
    \label{fig:forces}
\end{figure}

\subsection{Intensity and pressure}
\label{sec:Int_FL}
\begin{figure}
    \centering
    \includegraphics[width=\linewidth]{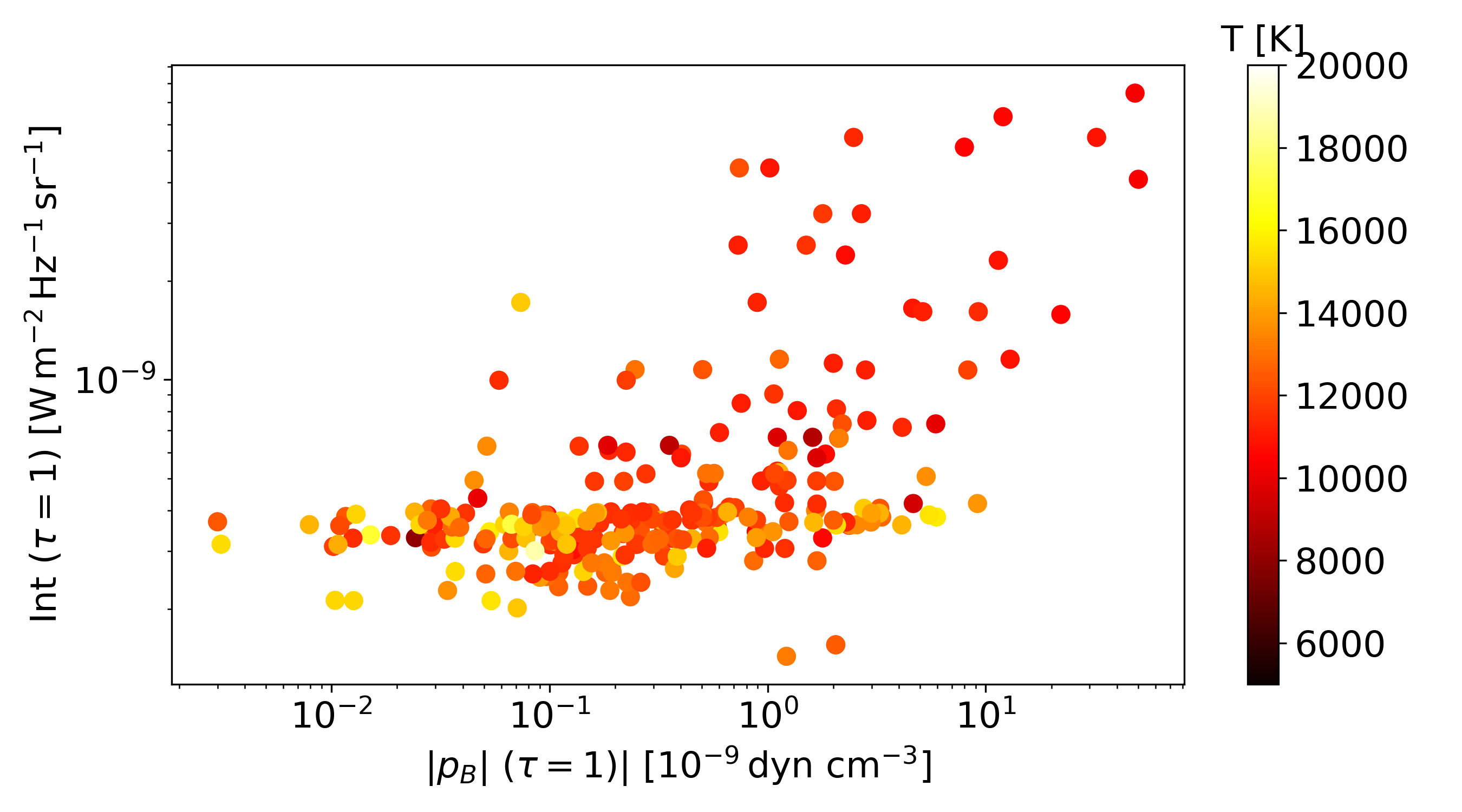}
    \caption{Correlation between the absolute value of magentic pressure and intensity at the apex of the coronal loop along our LOS (both parameters extracted at specific locations of $\tau=1$ corresponding to wavelengths $Q_1$, $Q_2$, and $Q_3$).}
    \label{fig:FLInt_relation}
\end{figure}

According to \cite{Wenzhi2024}, there is a balance in the system between magnetic pressure and gas pressure in the vertical direction at the apex of a flare loop. In the gradual phase, before any coronal rain forms, they show $(\textbf{J}\times\textbf{B}-\nabla p) \cdot \textbf{e}_z \approx 0$. The authors analysed this relation and commented that this balance is dynamic as it adjusts with time and cooling of the loop system throughout the simulation. In Fig.~\ref{fig:forces} we show the balance for the simulation presented here \citep[for consistency with][in the lower panel we plot the force balance on a linear scale, which results in a predominantly white panel]{Wenzhi2024}. The upper panel of the figure shows the corresponding density structure in the domain at that particular moment. Once the condensations form, they push the system out of its initial vertical equilibrium between the pressure gradient and the Lorentz force, and oscillations perpendicular to the magnetic field form \citep[see][]{Jercic2024}. Hence, unlike \cite{Wenzhi2024} we still needed to take the gravity force into account to achieve an equilibrium between forces. In the bottom panel of Fig.~\ref{fig:forces}, we show $F_{tot}$, which marks the total force (vertical component of Lorentz + pressure gradient force + gravity force). It is predominantly zero, except at locations of falling (or rising) threads. If we only restrict ourselves at the apex of the loop, gravity represents a sort of transient. If we were to average in time over a given period, gravity would cancel out, and hence it is not the dominant factor here. According to \cite{Wenzhi2024}, we should be able to estimate the magnetic field by knowing the thermal pressure due to this thermal and magnetic pressure balance at the apex of the loop. \cite{Sahin2024} used this relation to estimate the magnetic field from their observations, and they obtained very consistent results when comparing the loss in magnetic energy with the thermal energy in a flare. Considering the results of Sect.~\ref{sec:Correlations}, this means that, through pressure and intensity correlation, there should be a relation between intensity and magnetic-field strength. To check for this, we plotted values of intensity and the magnetic pressure for every $Q_1$, $Q_2$, and $Q_3$ wavelength at their corresponding $\tau=1$ locations. Our LOS is relatively low, and because Mg~II is an optically thick line we do not 'see' the threads at the apex of the loop, but we tend to capture emission from the falling threads where the balance between the thermal and magnetic pressure is lost. Fitting these data  would result in a large scatter of data. To avoid this, we limited our LOS in the range of $x=-10$ to 10\,Mm in order to only capture the apex of the low-lying loops near the centre of the LOS. This allowed us to confirm if the magnetic pressure and intensity are related. In Fig.~\ref{fig:FLInt_relation} we plot a logarithmically scaled relationship of intensity and the |$p_B$| at the apex of the coronal loop. The Pearson correlation coefficient between the two parameters is 0.65, which is a good indicator of an existing correlation.

\subsection{Mg~II triplet lines}
Significant amounts of work concerning the analysis of Mg~II triplet lines has been predominantly focused on the chromosphere. We give a more detailed account of such emission for cold coronal condensations above the limb. 

\begin{figure}
    \centering
    \includegraphics[width=0.8\linewidth]{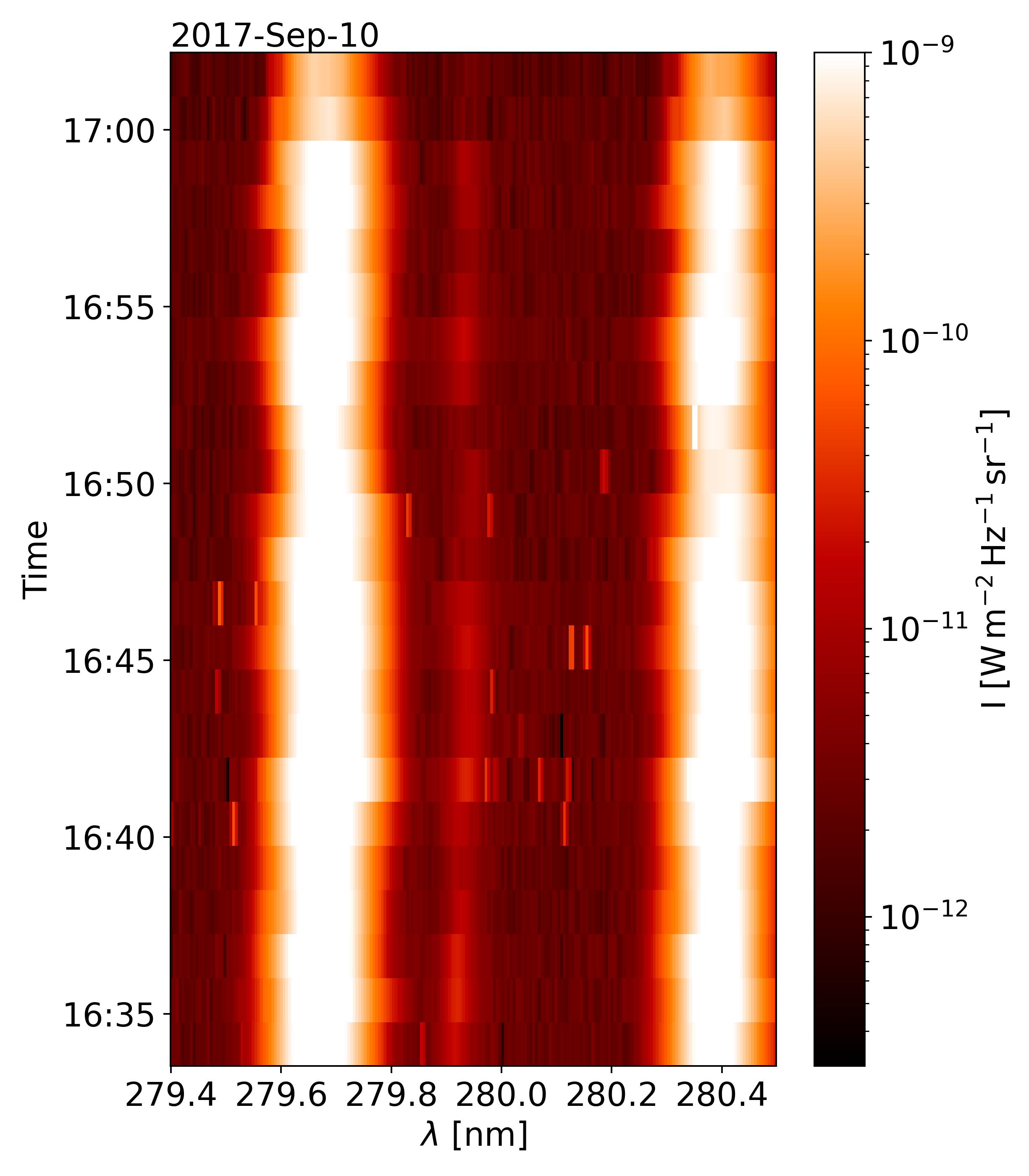}
    \caption{Average emission of blended subordinate lines (2798.75 and 2798.82\,\AA) located between Mg~II~k and h lines for September 2017. We averaged part of the first slit in the raster from the 100th to the 120th pixel (from 951.43 to 957.23\,arcsec) high above the chromosphere, during which there is a clear, substantially increased emission in the subordinate lines (16:34:09 until 17:00:19).}
    \label{fig:sub_emission_obs}
\end{figure}

\begin{figure}
    \centering
    \includegraphics[width=0.8\linewidth]{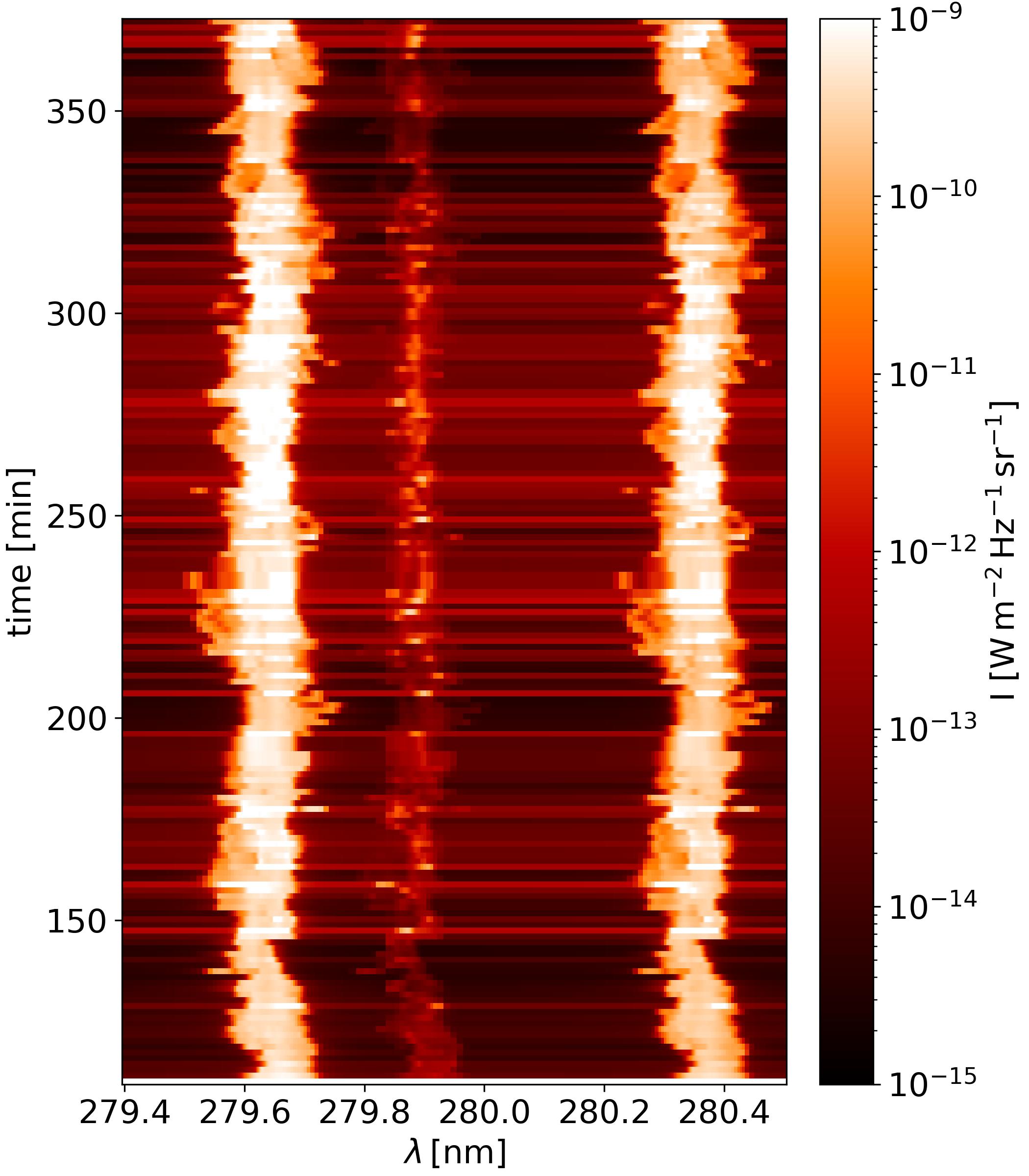}
    \caption{Synthetic spectra showing the emission of blended subordinate lines (2798.75 and 2798.82\,\AA) located between Mg~II~k and h lines. We show the full period of the simulation after the condensations form (more than 4\,hr) for the LOS cut at a 10\,Mm height.}
    \label{fig:sub_emission_sim}
\end{figure}

We observe an increased line emission for a limited time and area in the September 2017 observations in the 2798.75 and 2798.82\,\AA~lines (blended together, as seen in Fig.~\ref{fig:sub_emission_obs}). From here on, we only discuss these two blended lines that we can observe, and we continue referring to them as the triplet lines. Figure~\ref{fig:sub_emission_obs} shows the log plot of intensity and its evolution in time for the September 2017 case. To have physical units and to be able to make a comparison with the simulation, we performed radiometric calibration and PSF deconvolution using \texttt{irispreppy} package \footnote{Available at \url{https://github.com/OfAaron3/irispreppy} or via pip.} \citep{Courrier2018}. We show here the average intensity for pixels 100--120 (from 951.43" to 957.23") for the first slit in the raster. This averaging allowed us to show that there exists a clear, substantially increased emission in the triplet lines (16:34:09 until 17:00:19 UT). Other slits also show this increased emission in the triplet lines; however, the exact location and the intensity above the background all vary with time. We chose to show only the first slit simply because it was the first one to show this increase. For the observations of December 2021 and June 2017 we did not detect a similar increase in emission of the triplet lines. Figure~\ref{fig:sub_emission_sim} shows the same as Fig.~\ref{fig:sub_emission_obs} but for the case of the simulation. The figure does not show the average over multiple pixels, rather it depicts the sequential evolution in time of the Mg~II spectral line  with a LOS at $y=10$\,Mm from the moment there is condensation in the simulation.
 Figures~\ref{fig:sub_emission_obs} and~\ref{fig:sub_emission_sim} are directly comparable. We plot the same range of wavelengths and the same maximum value for this reason. We kept the maximum of the colour bar at a particular value of 10$^{-9}$\,W\,m$^{-2}$\,Hz$^{-1}$\,sr$^{-1}$ in order to better see the triplet lines. In the case of observations, both Mg~II~h and~k, as well as the triplet lines, lose their finer variations due to averaging (as seen in Fig.~\ref{fig:sub_emission_obs}). Nonetheless, the emission of the triplet lines is very obvious and the two figures show closely comparable values.

\section{Discussion}
\label{sec:discors}
We started off with the aim of improving the diagnostic capabilities that would allow us to disentangle the complex atmospheres at the source of the observed spectra of solar prominences. In this section we discuss the results and how we can leverage this simulation to relate the observables to the atmospheric parameters of the simulation. We can thus use the information we inferred to interpret the observations. 

\subsection{Quartile analysis}
The parameters derived from $Q_1$, $Q_2,$ and $Q_3$ (width, asymmetry, and the Doppler shift) provide a systematic way of analysing both observations and simulation. The location where different parts of the spectra ($Q_1$, $Q_2$, and $Q_3$) form can significantly vary during the evolution of the threads. They do not necessarily all form in the same thread and can be considerably spread out along the LOS (e.g. Fig.~\ref{fig:q123_locations}). The factor determining where the different parts of the spectra will form is an interplay of opacity and velocity of the threads along the LOS. These parameters are highly intricate, and choosing only one representative pixel allowed us to perform a more direct comparison between observations and simulation.

Each observation captures falling threads with a different LOS with respect to their direction of motion. We see that December 2021 exhibits significantly smaller variations than September 2017 and the simulation. As these two have the most comparable LOS with respect to the motion of the threads, they both show a similar range of values for width, asymmetry, and Doppler shift. June 2017, on the other hand, captures threads that are vertically falling at the location where we performed the analysis. Hence, the LOS is capturing the motion perpendicular to the LOS rather than along it, and the changes we see for June 2017 are closer to values of December 2021 than September 2017 and the simulation. The changes in width and asymmetry both point to a similar effect, the motion of different threads along the LOS. A larger width indicates that there are more different structures moving in opposite directions (a sort of counter-streaming motion). The LOS in the case of the simulation and September 2017 is along the motion and captures this, while for December 2021 and June 2017 the LOS is perpendicular to the motion and cannot capture the variations as well. Although, this does not necessarily mean they are not there. We calculated the POS velocity for December 2021 by following a bright structure and approximating its fall with a straight line. For two very clear and bright structures the calculated POS velocities are $\sim$24 and $\sim$97\,km\,s$^{-1}$, which gives a range of values in better agreement with the Doppler shift in the simulation and the September 2017 observation.

If we compare the values of width and asymmetry to other works that applied this method to other coronal condensations (prominences), we measure similar values. For example, the study of \cite{Ruan2018} compared values calculated using the quartile (12, 50, and 88\% of the cumulative distribution of intensity) and the Gaussian method. Their range of values (0.2--0.8~\AA) for the most part matches with the observation of September 2017 and the simulation, although they show a larger peak value than what we measured for the specific slits we chose here. For the asymmetry, the values they measured (-0.52 to 0.21) fall into the range of values shown here. \cite{Peat2024_observations} also found similar values. Similarly to \cite{Ruan2018} they used the 12, 50, and 88\% quartiles. Their figures show values of width predominantly in the range of 0.25--0.5\,\AA~and asymmetry between -0.5 and +0.5. Considering both of these works used different quartiles (12 and 88 rather than 25 and 75\%) some difference in the results of width and asymmetry is to be expected. It is important, however, to keep in mind that the condensations presented in these two papers are of a completely different type than the coronal rain we studied here. Hence, the match in the results merely gives a confirmation of the method and the fact that the structures show similar values of dynamically related parameters. 

We see that the best match in the quartile parameters is between the simulation and the September 2017 case. For the other two observations the comparison is more complex due to a different LOS with respect to the motion of the rain. Additionally, if we chose data of a different pixel along a slit (or a different slit), we would obtain slightly different variations in the analysed parameters. Quartile analysis provides a general idea of the behaviour of different parameters describing the spectra, but alone it is not sufficient for a detailed comparison and strong conclusions on how well the observations match the simulation.

\subsection{Correlations}
There have been other studies looking at how exactly different observables relate to atmospheric parameters of the model atmosphere. \cite{Heinzel2014} looked at 1D isobaric slab models with different temperature distributions (isothermal, including a CCTR and in radiative equilibrium). Their models considered H and Mg~II lines. We also mention \cite{Levens2019}; the authors extended the parameter grid of \cite{Heinzel2014} (also static 1D slab models, which are isothermal-isobaric or include a CCTR). Both of these studies used observational values for the incident radiation (although each used different observations); nonetheless, their results are qualitatively similar \citep{Levens2019}. \cite{Heinzel2014} found that low pressure results in Mg~II lines without central reversal, while for higher pressures, central reversal increases with higher temperatures. Besides that, their line width is influenced by the gas pressure and the thickness of the 1D slab. \cite{Levens2019} found that intensities generally increase with higher pressures and slab thicknesses. They concluded that the greater intensities with higher pressures are related to collisional processes beginning to dominate over the radiative ones. Using \texttt{Promweaver,} we can confirm this conclusion. We indeed see a trend of higher intensities corresponding to locations of greater density (and pressure), which have collisional rates dominating over the radiative ones. Keeping in mind the already described individual complexity of the threads, there exists a scatter to this trend similar to the scatter seen in Fig.~\ref{fig:correlations}. 

Both \cite{Levens2019} and \cite{Heinzel2014} found a correlation of intensity with temperature. With increasing temperature the integrated intensity initially increases, and then it drops from about 10 to 40~000\,K \citep[see Fig.~8 in][]{Levens2019}. The authors explain this decrease in intensity with high enough temperatures as the result of the ionisation of Mg~II to Mg~III. The range of temperatures in the simulation presented here does not reach such high values (at the explored $\tau=1$ locations). However, even for the temperatures we analysed (in the range of $5000-20~000$\,K) we find no relation between temperature and intensity, while we confirm a strong relation between intensity and pressure and show that there exists a correlation between intensity and density. Moreover, there is a relation between the k/h ratio and the logarithm of pressure. \cite{Levens2019} also commented on this relation; however, they find that the k/h ratio does not depend strongly on the pressure. In their models that include a CCTR, they find some effect of the central pressure (at the centre of the slab) on the k/h ratio (the authors did not report the exact values of their correlation coefficients). When comparing our results with those of \cite{Levens2019} we need to keep in mind that we focused on the $\tau=1$ layer and the pressure there. This layer might not always reflect the central pressure \citep[considered in][]{Levens2019}, and hence we did not necessarily compare exactly the same pressures. In our case we find a moderate correlation of the $logp$ on the k/h ratio, with a Pearson correlation coefficient of 0.6. This is expected considering that both k- and h-line integrated intensities scale linearly with pressure in $Q_2$ (with Pearson correlation coefficients of 0.74 and 0.66) with different slopes. Because of this, their ratio increases with pressure in a roughly logarithmic manner. 

The 1D slab models represent the first (and essential) step needed to start disentangling the complex and intricate relation between the atmospheric parameters and the resulting spectral lines we see in observations. The next step comes in higher dimensionality models, with which comes more complex physics. We interpret the differences we see in the here presented spectra resulting from the 2.5D simulation versus the ones resulting from the 1D slab model of \cite{Levens2019} as a result of limitations of these 1D models in comparison to the 2.5D one. The first limitation is the assumption of an isothermal isobaric slab, for which we know prominences and coronal rain do not follow. Secondly, even the more complex 1D slab with a CCTR has a relatively simple transition, which lacks the complexity of the self-consistently formed thread we see in the 2.5D model (Fig.~\ref{fig:4pp1}). The conclusions of both \cite{Levens2019} and \cite{Heinzel2014} emphasise the importance of including the CCTR in the calculations. Sect.~\ref{sec:4pp} and Figs.~\ref{fig:4pp1} and~\ref{fig:4pp2} show the complexity of a single thread in great detail and, even more so, of the full domain once there are many such threads along the LOS. The fact that our simulation self-consistently creates these condensations results in conditions far from LTE, we are dealing with a more complex temperature and pressure structure than usually modelled in 1D or even 2D slab models. Unlike the models of \cite{Heinzel2014} and \cite{Levens2019}, calculations here take velocity into account. The velocity of the threads additionally influences the already complex pressure and density distribution resulting from the particular formation process. The system is highly non-linear and results in convoluted relations between parameters describing the source structures (the threads) and the parameters characterising the resulting spectral line (observables).

\subsection{Intensity and Lorentz force}
We show that the intensity and thermal pressure relation we find can be used to relate intensity and the magnetic pressure. Due to the particular LOS we chose, gravity also plays a role, and for the threads closer to the coronal-loop foot points we cannot claim there exists a good enough balance between the magnetic and thermal pressure. To still be able to analyse this relationship, we limited our LOS to the apex of the loop. This is why for the purposes of Fig.~\ref{fig:FLInt_relation} we cut the \texttt{AMRVAC} domain to only $x=-10$ to 10\,Mm. The result is shown in Fig.~\ref{fig:FLInt_relation} and demonstrates a relationship. A Pearson correlation coefficient of 0.65 even indicates a strong relationship. Nonetheless, fitting this relationship proved to be quite challenging. None of the fitting functions we attempted to use (line, quadratic, cubic, power law) managed to capture the behaviour of the points located between 4$\times10^{-11}$ and 2$\times10^{-9}$\,dyn\,cm$^{-3}$. The scatter of the particular dataset is large, and all the functions we tried similarly struggle to fully capture the behaviour between the magnitude of the gradient of magnetic pressure and intensity. Still, this relationship suggests that the intensity may be correlated with the magnetic field. However, we do not find such a direct correlation. Because our LOS exhibits many strong variations due to the condensations with starkly different atmospheric properties than the ambient corona, the gradient of the squared magnetic field and the magnetic field itself are different enough that a correlation with one does not imply a correlation with the other.

\subsection{Mg~II triplet lines}
Lastly, we take a closer look into the emission of Mg~II triplet lines in the off-limb condensation, in the simulation as well as in the observations. We see a substantially increased emission in the triplet lines found in the blue wing of Mg~II~k line. We only see this in the observation of September 2017, and only for a period of about 25\,min after the condensation formed. In the simulation, the increased emission of these triplet lines is continuous. This can be easily explained by the continuous influx of energy with our stochastic heating that keeps adding energy. In contrast, in the real flare system, after the sudden release of energy that energy is transformed into other forms \citep{Shibata_Magara2011} and dissipates with time.

The stochastic heating we imposed in the simulation results in the events similar to those observed in the aftermath of an X-class flare (September 2017). Figures~\ref{fig:sub_emission_obs} and ~\ref{fig:sub_emission_sim} show intensity for the September 2017 case and the simulation, respectively. As both figures have the same maximum value of the colour bar it is possible to directly compare them. We see the same type of increased emission in the simulation as well as in the observation. To further compare this, we consider the heating function given in Eq.~\ref{eq:Hi} and calculate the influx of energy from the pulses. Assuming the maximum value of this heating ($x=x_i$, $y=y_i$ and $t=t_i+\delta t_i/2)$, for a single event we obtain a value of $H_i = 0.33$\,erg\,cm$^{-3}$\,s$^{-1}$. As the spatial extent of a pulse is 2\,Mm (in $x$ and $y$) we can assume an influx of 6.6$\times$10$^7$\,erg\,cm$^{-2}$\,s$^{-1}$. Comparing this with the values in Table 1 of \cite{Withbroe_Noyes1977} we see that this range of values matches the one expected for an active region. Furthermore, outside of the condensations, the corona in the simulation matches the temperatures one would expect an active region to have. After the coronal rain forms, the corona reaches maximum temperatures in the range of 4.5 to $\sim$7\,MK. Overall, this simulation points to extreme conditions as a result of strong stochastic heating.

\section{Conclusions}
\label{sec:conclusion}
We used a 2.5D simulation of coronal condensations \citep{Jercic2024} in conjunction with the \texttt{Promweaver} framework and created synthetic spectra of Mg~II~h and k lines. We compared those spectra to observations of IRIS. The observations we chose are examples of flare-driven coronal rain and a quiescent coronal rain above an active region.

Analysing the contribution function of intensity for two particular moments in the simulation (just after the formation and when the domain is filled with threads) we see that the opacity of different structures along the LOS and their velocity (Doppler shift) influence how the spectra will look and which threads will be responsible for its different parts. Early on in the simulation we see how the source function changes in a single thread (Fig.~\ref{fig:4pp1}). This demonstrates that even a single thread presents an example of a complex pressure and density structure with a varying source function. Looking at the later moment, the situation becomes considerably more complex, with different sections along the LOS contributing to the resulting spectral line (see Fig.~\ref{fig:q123_locations}). 

The simulation shows thread-like condensations first appearing lower down in the atmosphere. With time, these condensations start appearing higher up and eventually fill out most of the domain. This is similar to how the observed flare-driven coronal rain first appears at lower heights and only gradually reaches higher heights, as longer loops need more time to cool (as can be seen in the time sequence in Fig.~\ref{fig:iris_Dec2021}). With the quartile analysis we obtain an overview of how different parameters change in time, and we see there is a good match in width and asymmetry, as well as in the Doppler shift values of the simulation with the September 2017 data. The two have the same LOS with respect to the coronal rain motion, in contrast to the other datasets we considered. This indicates that point of view is an important aspect to consider when analysing such parameters. Furthermore, similarly to the September 2017 case, the simulation shows an increased emission in the Mg~II triplet lines. This indicates that the strong stochastic heating imposed in the simulation is capable of creating similar extreme conditions of a solar active region. We demonstrate an important relationship between the magnetic and thermal pressure. Despite intensity being correlated with both those quantities, we find that this does not suffice to conclude the same correlation with the magnetic field. As the condensations represent strong variations along the LOS, the influence of the gradient is significant enough that the magnetic field itself does not show a similar trend to the magnetic pressure.

The current work shows that the 2.5D simulation matches the observations, and it matches that of a flare-driven coronal rain particularly well. We successfully recreated a similar evolution of the spectral parameters as well as the observed spectral shape and values (Fig.~\ref{fig:quartile_analysis}). Due to the greater complexity of a 2.5D simulation (in comparison to 1D slab models), the results here show convoluted relations between different parameters describing the spectra and those relating to the simulated atmosphere. Nonetheless, we still recover certain strong correlations that will be of use to observers (e.g. intensity and density). On the other hand, this complexity of the system also puts some expected correlations into question (e.g. intensity and magnetic field). This research provides new insights into the simulation and observation of coronal rain, answering some questions while raising new ones (e.g. how exactly the emission of the triplet lines is related to the continuous localised heating and whether it can be used as its estimate). The simulation we analysed here is only 2.5D, and a caveat concerning the correlations in Fig.~\ref{fig:correlations} is that they were only measured for one specific LOS. This needs additional analysis, preferably with a 3D simulation, so that we can explore different LOSs with different angles with respect to the motion of the condensations. Furthermore, working in only 2.5D can limit our ability to simulate certain processes (e.g. a reconnection event) and in such a way influence the results. As a follow up we plan to create a 3D simulation, analyse its dynamics, and study how it differs as we look at it from different angles. Moreover, \texttt{Promweaver} is a 1.5D code, and a multi-dimensional version of it now exists: \texttt{DexRT} \citep{Osborne2025}. In the future we also plan to utilise \texttt{DexRT} and explore the differences between the two codes.

\section*{Data availability}
The scripts needed to recreate the dataset (and plots) are easily available in the Zenodo repository, \url{https://doi.org/10.5281/zenodo.21441461}. Due to the size constraints, the dataset used in this study is available on request.

\begin{acknowledgements}
       We thank the referee for the feedback that helped improve the clarity of the paper. Dr. Jer\v{c}i\'{c}'s research was made possible by an appointment to the NASA Postdoctoral Program at the Goddard Space Flight Center, administered by Oak Ridge Associated Universities under contract with NASA. A.~G.~M.~P. was supported by grant PI~2102/1-1 from the Deutsche Forschungsgemeinschaft (DFG). J.~M.~J. acknowledges support through the European Space Agency (ESA) Research Fellowship Program in Space Science. IRIS is a NASA small explorer mission developed and operated by LMSAL with mission operations executed at NASA Ames Research Center and major contributions to downlink communications funded by ESA and the Norwegian Space Centre.
\end{acknowledgements}

  \bibliographystyle{aa} 
  \bibliography{references} 

@ARTICLE{Antiochos_Klimchuk1991,
       author = {{Antiochos}, S.~K. and {Klimchuk}, J.~A.},
        title = "{A Model for the Formation of Solar Prominences}",
      journal = {\apj},
         year = 1991,
        month = sep,
       volume = {378},
        pages = {372},
          doi = {10.1086/170437},
       adsurl = {https://ui.adsabs.harvard.edu/abs/1991ApJ...378..372A}
}

@ARTICLE{Antolin2012,
       author = {{Antolin}, Patrick and {Vissers}, Gregal and {Rouppe van der Voort}, Luc},
        title = "{On-Disk Coronal Rain}",
      journal = {\solphys},
         year = 2012,
        month = oct,
       volume = {280},
       number = {2},
        pages = {457-474},
          doi = {10.1007/s11207-012-9979-7},
archivePrefix = {arXiv},
       eprint = {1203.2077},
 primaryClass = {astro-ph.SR},
       adsurl = {https://ui.adsabs.harvard.edu/abs/2012SoPh..280..457A}
}

@ARTICLE{Antolin2015,
       author = {{Antolin}, P. and {Vissers}, G. and {Pereira}, T.~M.~D. and {Rouppe van der Voort}, L. and {Scullion}, E.},
        title = "{The Multithermal and Multi-stranded Nature of Coronal Rain}",
      journal = {\apj},
         year = 2015,
        month = jun,
       volume = {806},
       number = {1},
          eid = {81},
        pages = {81},
          doi = {10.1088/0004-637X/806/1/81},
archivePrefix = {arXiv},
       eprint = {1504.04418},
 primaryClass = {astro-ph.SR},
       adsurl = {https://ui.adsabs.harvard.edu/abs/2015ApJ...806...81A}
}

@ARTICLE{Antolin2020,
       author = {{Antolin}, Patrick},
        title = "{Thermal instability and non-equilibrium in solar coronal loops: from coronal rain to long-period intensity pulsations}",
      journal = {Plasma Physics and Controlled Fusion},
         year = 2020,
        month = jan,
       volume = {62},
       number = {1},
          eid = {014016},
        pages = {014016},
          doi = {10.1088/1361-6587/ab5406},
       adsurl = {https://ui.adsabs.harvard.edu/abs/2020PPCF...62a4016A}
}

@ARTICLE{Antolin_Froment2022,
       author = {{Antolin}, Patrick and {Froment}, Clara},
        title = "{Multi-Scale Variability of Coronal Loops Set by Thermal Non-Equilibrium and Instability as a Probe for Coronal Heating}",
      journal = {Frontiers in Astronomy and Space Sciences},
         year = 2022,
        month = mar,
       volume = {9},
          eid = {820116},
        pages = {820116},
          doi = {10.3389/fspas.2022.820116},
       adsurl = {https://ui.adsabs.harvard.edu/abs/2022FrASS...920116A}
}

@ARTICLE{Auer1994,
       author = {{Auer}, L.~H. and {Paletou}, F.},
        title = "{Two-dimensional radiative transfer with partial frequency redistribution I. General method}",
      journal = {\aap},
         year = 1994,
        month = may,
       volume = {285},
        pages = {675-686},
       adsurl = {https://ui.adsabs.harvard.edu/abs/1994A&A...285..675A}
}

@ARTICLE{Ballester2018,
       author = {{Ballester}, Jos{\'e} Luis and {Alexeev}, Igor and {Collados}, Manuel and {Downes}, Turlough and {Pfaff}, Robert F. and {Gilbert}, Holly and {Khodachenko}, Maxim and {Khomenko}, Elena and {Shaikhislamov}, Ildar F. and {Soler}, Roberto and {V{\'a}zquez-Semadeni}, Enrique and {Zaqarashvili}, Teimuraz},
        title = "{Partially Ionized Plasmas in Astrophysics}",
      journal = {\ssr},
         year = 2018,
        month = mar,
       volume = {214},
       number = {2},
          eid = {58},
        pages = {58},
          doi = {10.1007/s11214-018-0485-6},
archivePrefix = {arXiv},
       eprint = {1707.07975},
 primaryClass = {astro-ph.SR},
       adsurl = {https://ui.adsabs.harvard.edu/abs/2018SSRv..214...58B}
}

@ARTICLE{Carlsson_Stein1997,
       author = {{Carlsson}, Mats and {Stein}, Robert F.},
        title = "{Formation of Solar Calcium H and K Bright Grains}",
      journal = {\apj},
         year = 1997,
        month = may,
       volume = {481},
       number = {1},
        pages = {500-514},
          doi = {10.1086/304043},
       adsurl = {https://ui.adsabs.harvard.edu/abs/1997ApJ...481..500C}
}

@ARTICLE{Courrier2018,
       author = {{Courrier}, Hans and {Kankelborg}, Charles and {De Pontieu}, Bart and {W{\"u}lser}, Jean-Pierre},
        title = "{An on Orbit Determination of Point Spread Functions for the Interface Region Imaging Spectrograph}",
      journal = {\solphys},
         year = 2018,
        month = sep,
       volume = {293},
       number = {9},
          eid = {125},
        pages = {125},
          doi = {10.1007/s11207-018-1347-9},
       adsurl = {https://ui.adsabs.harvard.edu/abs/2018SoPh..293..125C}
}

@ARTICLE{DePontieu2014,
       author = {{De Pontieu}, B. and {Title}, A.~M. and {Lemen}, J.~R. and {Kushner}, G.~D. and {Akin}, D.~J. and {Allard}, B. and {Berger}, T. and {Boerner}, P. and {Cheung}, M. and {Chou}, C. and {Drake}, J.~F. and {Duncan}, D.~W. and {Freeland}, S. and {Heyman}, G.~F. and {Hoffman}, C. and {Hurlburt}, N.~E. and {Lindgren}, R.~W. and {Mathur}, D. and {Rehse}, R. and {Sabolish}, D. and {Seguin}, R. and {Schrijver}, C.~J. and {Tarbell}, T.~D. and {W{\"u}lser}, J. -P. and {Wolfson}, C.~J. and {Yanari}, C. and {Mudge}, J. and {Nguyen-Phuc}, N. and {Timmons}, R. and {van Bezooijen}, R. and {Weingrod}, I. and {Brookner}, R. and {Butcher}, G. and {Dougherty}, B. and {Eder}, J. and {Knagenhjelm}, V. and {Larsen}, S. and {Mansir}, D. and {Phan}, L. and {Boyle}, P. and {Cheimets}, P.~N. and {DeLuca}, E.~E. and {Golub}, L. and {Gates}, R. and {Hertz}, E. and {McKillop}, S. and {Park}, S. and {Perry}, T. and {Podgorski}, W.~A. and {Reeves}, K. and {Saar}, S. and {Testa}, P. and {Tian}, H. and {Weber}, M. and {Dunn}, C. and {Eccles}, S. and {Jaeggli}, S.~A. and {Kankelborg}, C.~C. and {Mashburn}, K. and {Pust}, N. and {Springer}, L. and {Carvalho}, R. and {Kleint}, L. and {Marmie}, J. and {Mazmanian}, E. and {Pereira}, T.~M.~D. and {Sawyer}, S. and {Strong}, J. and {Worden}, S.~P. and {Carlsson}, M. and {Hansteen}, V.~H. and {Leenaarts}, J. and {Wiesmann}, M. and {Aloise}, J. and {Chu}, K. -C. and {Bush}, R.~I. and {Scherrer}, P.~H. and {Brekke}, P. and {Martinez-Sykora}, J. and {Lites}, B.~W. and {McIntosh}, S.~W. and {Uitenbroek}, H. and {Okamoto}, T.~J. and {Gummin}, M.~A. and {Auker}, G. and {Jerram}, P. and {Pool}, P. and {Waltham}, N.},
        title = "{The Interface Region Imaging Spectrograph (IRIS)}",
      journal = {\solphys},
         year = 2014,
        month = jul,
       volume = {289},
       number = {7},
        pages = {2733-2779},
          doi = {10.1007/s11207-014-0485-y},
archivePrefix = {arXiv},
       eprint = {1401.2491},
 primaryClass = {astro-ph.SR},
       adsurl = {https://ui.adsabs.harvard.edu/abs/2014SoPh..289.2733D}
}

@ARTICLE{De_Wilde2025,
       author = {{De Wilde}, M. and {Pietrow}, A.~G.~M. and {Druett}, M.~K. and {Pastor Yabar}, A. and {Koza}, J. and {Kontogiannis}, I. and {Andriienko}, O. and {Berlicki}, A. and {Brunvoll}, A.~R. and {de la Cruz Rodr{\'\i}guez}, J. and {Thoen Faber}, J. and {Joshi}, R. and {Kuridze}, D. and {N{\'o}brega-Siverio}, D. and {Rouppe van der Voort}, L.~H.~M. and {Ryb{\'a}k}, J. and {Scullion}, E. and {Silva}, A.~M. and {Vashalomidze}, Z. and {Vicente Ar{\'e}valo}, A. and {Wi{\'s}niewska}, A. and {Yadav}, R. and {Zaqarashvili}, T.~V. and {Zbinden}, J. and {{\O}yre}, E.~S.},
        title = "{Synthesizing Sun-as-a-star flare spectra from high-resolution solar observations}",
      journal = {\aap},
         year = 2025,
        month = aug,
       volume = {700},
          eid = {A275},
        pages = {A275},
          doi = {10.1051/0004-6361/202554870},
archivePrefix = {arXiv},
       eprint = {2507.07967},
 primaryClass = {astro-ph.SR},
       adsurl = {https://ui.adsabs.harvard.edu/abs/2025A&A...700A.275D}
}

@ARTICLE{Fontenla1993,
       author = {{Fontenla}, J.~M. and {Avrett}, E.~H. and {Loeser}, R.},
        title = "{Energy Balance in the Solar Transition Region. III. Helium Emission in Hydrostatic, Constant-Abundance Models with Diffusion}",
      journal = {\apj},
         year = 1993,
        month = mar,
       volume = {406},
        pages = {319},
          doi = {10.1086/172443},
       adsurl = {https://ui.adsabs.harvard.edu/abs/1993ApJ...406..319F}
}

@ARTICLE{Gunar2008,
       author = {{Gun{\'a}r}, S. and {Heinzel}, P. and {Anzer}, U. and {Schmieder}, B.},
        title = "{On Lyman-line asymmetries in quiescent prominences}",
      journal = {\aap},
         year = 2008,
        month = oct,
       volume = {490},
       number = {1},
        pages = {307-313},
          doi = {10.1051/0004-6361:200810127},
       adsurl = {https://ui.adsabs.harvard.edu/abs/2008A&A...490..307G}
}

@ARTICLE{Heasley_Kneer1976,
       author = {{Heasley}, J.~N. and {Kneer}, F.},
        title = "{Formation of spectral lines with partial frequency redistribution.}",
      journal = {\apj},
         year = 1976,
        month = feb,
       volume = {203},
        pages = {660-666},
          doi = {10.1086/154124},
       adsurl = {https://ui.adsabs.harvard.edu/abs/1976ApJ...203..660H}
}

@ARTICLE{Heinzel1987,
       author = {{Heinzel}, P. and {Gouttebroze}, P. and {Vial}, J. -C.},
        title = "{Formation of the hydrogen spectrum in quiescent prominences - One-dimensional models with standard partial redistribution}",
      journal = {\aap},
         year = 1987,
        month = sep,
       volume = {183},
       number = {2},
        pages = {351-362},
       adsurl = {https://ui.adsabs.harvard.edu/abs/1987A&A...183..351H}
}

@ARTICLE{Heinzel2001,
       author = {{Heinzel}, P. and {Anzer}, U.},
        title = "{Prominence fine structures in a magnetic equilibrium: Two-dimensional models with multilevel radiative transfer}",
      journal = {\aap},
         year = 2001,
        month = sep,
       volume = {375},
        pages = {1082-1090},
          doi = {10.1051/0004-6361:20010926},
       adsurl = {https://ui.adsabs.harvard.edu/abs/2001A&A...375.1082H}
}

@ARTICLE{Heinzel2014,
       author = {{Heinzel}, P. and {Vial}, J. -C. and {Anzer}, U.},
        title = "{On the formation of Mg ii h and k lines in solar prominences}",
      journal = {\aap},
         year = 2014,
        month = apr,
       volume = {564},
          eid = {A132},
        pages = {A132},
          doi = {10.1051/0004-6361/201322886},
       adsurl = {https://ui.adsabs.harvard.edu/abs/2014A&A...564A.132H}
}

@ARTICLE{Heinzel2015_letter,
       author = {{Heinzel}, P. and {Schmieder}, B. and {Mein}, N. and {Gun{\'a}r}, S.},
        title = "{Understanding the Mg II and H{\ensuremath{\alpha}} Spectra in a Highly Dynamical Solar Prominence}",
      journal = {\apjl},
         year = 2015,
        month = feb,
       volume = {800},
       number = {1},
          eid = {L13},
        pages = {L13},
          doi = {10.1088/2041-8205/800/1/L13},
       adsurl = {https://ui.adsabs.harvard.edu/abs/2015ApJ...800L..13H}
}

@ARTICLE{Heinzel2015,
       author = {{Heinzel}, P. and {Gun{\'a}r}, S. and {Anzer}, U.},
        title = "{Fast approximate radiative transfer method for visualizing the fine structure of prominences in the hydrogen H{\ensuremath{\alpha}} line}",
      journal = {\aap},
         year = 2015,
        month = jul,
       volume = {579},
          eid = {A16},
        pages = {A16},
          doi = {10.1051/0004-6361/201525716},
       adsurl = {https://ui.adsabs.harvard.edu/abs/2015A&A...579A..16H}
}

@INPROCEEDINGS{Heinzel_chapter2015,
       author = {{Heinzel}, Petr},
        title = "{Radiative Transfer in Solar Prominences}",
    booktitle = {Solar Prominences},
         year = 2015,
       editor = {{Vial}, Jean-Claude and {Engvold}, Oddbj{\o}rn},
       series = {Astrophysics and Space Science Library},
       volume = {415},
        month = jan,
        pages = {103},
          doi = {10.1007/978-3-319-10416-4_5},
       adsurl = {https://ui.adsabs.harvard.edu/abs/2015ASSL..415..103H}
}

@ARTICLE{Heinzel2025,
       author = {{Heinzel}, Petr and {Gun{\'a}r}, Stanislav},
        title = "{Radiative Transfer in Solar Prominences: An Overview and Current Trends}",
      journal = {\solphys},
         year = 2025,
        month = nov,
       volume = {300},
       number = {12},
          eid = {167},
        pages = {167},
          doi = {10.1007/s11207-025-02585-y},
       adsurl = {https://ui.adsabs.harvard.edu/abs/2025SoPh..300..167H}
}

@ARTICLE{Jenkins2022,
       author = {{Jenkins}, Jack M. and {Keppens}, Rony},
        title = "{Resolving the solar prominence/filament paradox using the magnetic Rayleigh-Taylor instability}",
      journal = {Nature Astronomy},
         year = 2022,
        month = jul,
       volume = {6},
        pages = {942-950},
          doi = {10.1038/s41550-022-01705-z},
       adsurl = {https://ui.adsabs.harvard.edu/abs/2022NatAs...6..942J}
}

@ARTICLE{Jenkins2023,
       author = {{Jenkins}, J.~M. and {Osborne}, C.~M.~J. and {Keppens}, R.},
        title = "{1.5D non-LTE spectral synthesis of a 3D filament and prominence simulation}",
      journal = {\aap},
         year = 2023,
        month = feb,
       volume = {670},
          eid = {A179},
        pages = {A179},
          doi = {10.1051/0004-6361/202244868},
archivePrefix = {arXiv},
       eprint = {2211.14869},
 primaryClass = {astro-ph.SR},
       adsurl = {https://ui.adsabs.harvard.edu/abs/2023A&A...670A.179J}
}

@ARTICLE{Jercic2023,
       author = {{Jer{\v{c}}i{\'c}}, V. and {Keppens}, R.},
        title = "{Dynamic formation of multi-threaded prominences in arcade configurations}",
      journal = {\aap},
         year = 2023,
        month = feb,
       volume = {670},
          eid = {A64},
        pages = {A64},
          doi = {10.1051/0004-6361/202245067},
archivePrefix = {arXiv},
       eprint = {2212.08537},
 primaryClass = {astro-ph.SR},
       adsurl = {https://ui.adsabs.harvard.edu/abs/2023A&A...670A..64J}
}

@ARTICLE{Jercic2024,
       author = {{Jer{\v{c}}i{\'c}}, V. and {Jenkins}, J.~M. and {Keppens}, R.},
        title = "{Prominence and coronal rain formation by steady versus stochastic heating and how we can relate it to observations}",
      journal = {\aap},
         year = 2024,
        month = aug,
       volume = {688},
          eid = {A145},
        pages = {A145},
          doi = {10.1051/0004-6361/202348442},
archivePrefix = {arXiv},
       eprint = {2406.02955},
 primaryClass = {astro-ph.SR},
       adsurl = {https://ui.adsabs.harvard.edu/abs/2024A&A...688A.145J}
}

@ARTICLE{Jercic2025,
       author = {{Jer{\v{c}}i{\'c}}, V. and {Popescu Braileanu}, B. and {Keppens}, R.},
        title = "{Forming Prominences Accounting for Partial Ionization Effects}",
      journal = {\apj},
         year = 2025,
        month = jun,
       volume = {986},
       number = {2},
          eid = {134},
        pages = {134},
          doi = {10.3847/1538-4357/add6aa},
archivePrefix = {arXiv},
       eprint = {2505.07990},
 primaryClass = {astro-ph.SR},
       adsurl = {https://ui.adsabs.harvard.edu/abs/2025ApJ...986..134J}
}

@ARTICLE{Karpen2001,
       author = {{Karpen}, J.~T. and {Antiochos}, S.~K. and {Hohensee}, M. and {Klimchuk}, J.~A. and {MacNeice}, P.~J.},
        title = "{Are Magnetic Dips Necessary for Prominence Formation?}",
      journal = {\apjl},
         year = 2001,
        month = may,
       volume = {553},
       number = {1},
        pages = {L85-L88},
          doi = {10.1086/320497},
       adsurl = {https://ui.adsabs.harvard.edu/abs/2001ApJ...553L..85K}
}

@ARTICLE{Keppens2020,
title = {MPI-AMRVAC: A parallel, grid-adaptive PDE toolkit},
journal = {Computers \& Mathematics with Applications},
volume = {81},
pages = {316-333},
year = {2021},
note = {Development and Application of Open-source Software for Problems with Numerical PDEs},
issn = {0898-1221},
doi = {https://doi.org/10.1016/j.camwa.2020.03.023},
url = {https://www.sciencedirect.com/science/article/pii/S0898122120301279},
author = {Rony Keppens and Jannis Teunissen and Chun Xia and Oliver Porth}
}

@ARTICLE{Keppens2023,
       author = {{Keppens}, R. and {Popescu Braileanu}, B. and {Zhou}, Y. and {Ruan}, W. and {Xia}, C. and {Guo}, Y. and {Claes}, N. and {Bacchini}, F.},
        title = "{MPI-AMRVAC 3.0: Updates to an open-source simulation framework}",
      journal = {\aap},
         year = 2023,
        month = may,
       volume = {673},
          eid = {A66},
        pages = {A66},
          doi = {10.1051/0004-6361/202245359},
archivePrefix = {arXiv},
       eprint = {2303.03026},
 primaryClass = {astro-ph.IM},
       adsurl = {https://ui.adsabs.harvard.edu/abs/2023A&A...673A..66K}
}

@ARTICLE{Kerr2015,
       author = {{Kerr}, G.~S. and {Sim{\~o}es}, P.~J.~A. and {Qiu}, J. and {Fletcher}, L.},
        title = "{IRIS observations of the Mg ii h and k lines during a solar flare}",
      journal = {\aap},
         year = 2015,
        month = oct,
       volume = {582},
          eid = {A50},
        pages = {A50},
          doi = {10.1051/0004-6361/201526128},
archivePrefix = {arXiv},
       eprint = {1508.03813},
 primaryClass = {astro-ph.SR},
       adsurl = {https://ui.adsabs.harvard.edu/abs/2015A&A...582A..50K}
}

@ARTICLE{Labrosse2010,
       author = {{Labrosse}, N. and {Heinzel}, P. and {Vial}, J. -C. and {Kucera}, T. and {Parenti}, S. and {Gun{\'a}r}, S. and {Schmieder}, B. and {Kilper}, G.},
        title = "{Physics of Solar Prominences: I{\textemdash}Spectral Diagnostics and Non-LTE Modelling}",
      journal = {\ssr},
         year = 2010,
        month = apr,
       volume = {151},
       number = {4},
        pages = {243-332},
          doi = {10.1007/s11214-010-9630-6},
archivePrefix = {arXiv},
       eprint = {1001.1620},
 primaryClass = {astro-ph.SR},
       adsurl = {https://ui.adsabs.harvard.edu/abs/2010SSRv..151..243L}
}

@ARTICLE{Lemaire_Gouttebroze1983,
       author = {{Lemaire}, P. and {Gouttebroze}, P.},
        title = "{Magnesium II line formation - The contribution of high atomic levels to the resonance lines}",
      journal = {\aap},
         year = 1983,
        month = sep,
       volume = {125},
       number = {2},
        pages = {241-245},
       adsurl = {https://ui.adsabs.harvard.edu/abs/1983A&A...125..241L}
}

@ARTICLE{Levens2019,
       author = {{Levens}, P.~J. and {Labrosse}, N.},
        title = "{Modelling of Mg II lines in solar prominences}",
      journal = {\aap},
         year = 2019,
        month = may,
       volume = {625},
          eid = {A30},
        pages = {A30},
          doi = {10.1051/0004-6361/201833132},
archivePrefix = {arXiv},
       eprint = {1902.00086},
 primaryClass = {astro-ph.SR},
       adsurl = {https://ui.adsabs.harvard.edu/abs/2019A&A...625A..30L}
}

@ARTICLE{Lin2005,
       author = {{Lin}, Yong and {Engvold}, Oddbj{\O}rn and {der Voort}, Luc Rouppe van and {Wiik}, Jun Elin and {Berger}, Thomas E.},
        title = "{Thin Threads of Solar Filaments}",
      journal = {\solphys},
         year = 2005,
        month = feb,
       volume = {226},
       number = {2},
        pages = {239-254},
          doi = {10.1007/s11207-005-6876-3},
       adsurl = {https://ui.adsabs.harvard.edu/abs/2005SoPh..226..239L}
}

@ARTICLE{Mackay2010,
       author = {{Mackay}, D.~H. and {Karpen}, J.~T. and {Ballester}, J.~L. and {Schmieder}, B. and {Aulanier}, G.},
        title = "{Physics of Solar Prominences: II{\textemdash}Magnetic Structure and Dynamics}",
      journal = {\ssr},
         year = 2010,
        month = apr,
       volume = {151},
       number = {4},
        pages = {333-399},
          doi = {10.1007/s11214-010-9628-0},
archivePrefix = {arXiv},
       eprint = {1001.1635},
 primaryClass = {astro-ph.SR},
       adsurl = {https://ui.adsabs.harvard.edu/abs/2010SSRv..151..333M}
}

@ARTICLE{Osborne2021,
       author = {{Osborne}, Christopher M.~J. and {Mili{\'c}}, Ivan},
        title = "{The Lightweaver Framework for Nonlocal Thermal Equilibrium Radiative Transfer in Python}",
      journal = {\apj},
         year = 2021,
        month = aug,
       volume = {917},
       number = {1},
          eid = {14},
        pages = {14},
          doi = {10.3847/1538-4357/ac02be},
archivePrefix = {arXiv},
       eprint = {2107.00475},
 primaryClass = {astro-ph.IM},
       adsurl = {https://ui.adsabs.harvard.edu/abs/2021ApJ...917...14O}
}

@ARTICLE{Osborne2025,
       author = {{Osborne}, C.~M.~J. and {Sannikov}, A.},
        title = "{Radiance cascades: a novel high-resolution formal solution for multidimensional non-LTE radiative transfer}",
      journal = {RAS Techniques and Instruments},
         year = 2025,
        month = jan,
       volume = {4},
          eid = {rzae062},
        pages = {rzae062},
          doi = {10.1093/rasti/rzae062},
archivePrefix = {arXiv},
       eprint = {2408.14425},
 primaryClass = {astro-ph.SR},
       adsurl = {https://ui.adsabs.harvard.edu/abs/2025RASTI...4...62O}
}

@ARTICLE{Paletou1993,
       author = {{Paletou}, F. and {Vial}, J. -C. and {Auer}, L.~H.},
        title = "{Two-dimensional radiative transfer with partial frequency redistribution. II. Application to resonance lines in quiescent prominences}",
      journal = {\aap},
         year = 1993,
        month = jul,
       volume = {274},
        pages = {571},
       adsurl = {https://ui.adsabs.harvard.edu/abs/1993A&A...274..571P}
}

@INPROCEEDINGS{Parenti2014,
       author = {{Parenti}, Susanna and {Vial}, Jean-Claude},
        title = "{On the nature of the prominence - corona transition region}",
    booktitle = {Nature of Prominences and their Role in Space Weather},
         year = 2014,
       editor = {{Schmieder}, Brigitte and {Malherbe}, Jean-Marie and {Wu}, S.~T.},
       series = {IAU Symposium},
       volume = {300},
        month = jan,
        pages = {69-78},
          doi = {10.1017/S1743921313010764},
       adsurl = {https://ui.adsabs.harvard.edu/abs/2014IAUS..300...69P}
}

@ARTICLE{Parenti2024,
       author = {{Parenti}, S. and {Luna}, M. and {Ballester}, J.~L.},
        title = "{Future prospects for partially ionized solar plasmas: the prominence case}",
      journal = {Philosophical Transactions of the Royal Society of London Series A},
         year = 2024,
        month = jun,
       volume = {382},
       number = {2272},
          eid = {20230225},
        pages = {20230225},
          doi = {10.1098/rsta.2023.0225},
archivePrefix = {arXiv},
       eprint = {2405.12627},
 primaryClass = {astro-ph.SR},
       adsurl = {https://ui.adsabs.harvard.edu/abs/2024RSPTA.38230225P}
}

@ARTICLE{Peat2023,
       author = {{Peat}, A.~W. and {Labrosse}, N. and {Gouttebroze}, P.},
        title = "{Mg II h\&k fine structure prominence modelling and the consequences for observations}",
      journal = {\aap},
         year = 2023,
        month = nov,
       volume = {679},
          eid = {A156},
        pages = {A156},
          doi = {10.1051/0004-6361/202347246},
archivePrefix = {arXiv},
       eprint = {2310.08249},
 primaryClass = {astro-ph.SR},
       adsurl = {https://ui.adsabs.harvard.edu/abs/2023A&A...679A.156P}
}

@ARTICLE{Peat2024_observations,
       author = {{Peat}, A.~W. and {Labrosse}, N. and {Barczynski}, K. and {Schmieder}, B.},
        title = "{Solar prominence diagnostics and their associated estimated errors from 1D NLTE Mg II h\&k modelling}",
      journal = {\aap},
         year = 2024,
        month = jun,
       volume = {686},
          eid = {A291},
        pages = {A291},
          doi = {10.1051/0004-6361/202348589},
archivePrefix = {arXiv},
       eprint = {2405.06492},
 primaryClass = {astro-ph.SR},
       adsurl = {https://ui.adsabs.harvard.edu/abs/2024A&A...686A.291P}
}

@ARTICLE{Pereira2015,
       author = {{Pereira}, Tiago M.~D. and {Carlsson}, Mats and {De Pontieu}, Bart and {Hansteen}, Viggo},
        title = "{The Formation of IRIS Diagnostics. IV. The Mg II Triplet Lines as a New Diagnostic for Lower Chromospheric Heating}",
      journal = {\apj},
         year = 2015,
        month = jun,
       volume = {806},
       number = {1},
          eid = {14},
        pages = {14},
          doi = {10.1088/0004-637X/806/1/14},
archivePrefix = {arXiv},
       eprint = {1504.01733},
 primaryClass = {astro-ph.SR},
       adsurl = {https://ui.adsabs.harvard.edu/abs/2015ApJ...806...14P}
}

@ARTICLE{Pietrow2024,
       author = {{Pietrow}, A.~G.~M. and {Liakh}, V. and {Osborne}, C.~M.~J. and {Jenkins}, J. and {Keppens}, R.},
        title = "{Spectral characteristics of a rotating solar prominence in multiple wavelengths}",
      journal = {\aap},
         year = 2024,
        month = oct,
       volume = {690},
          eid = {L15},
        pages = {L15},
          doi = {10.1051/0004-6361/202452276},
archivePrefix = {arXiv},
       eprint = {2410.03479},
 primaryClass = {astro-ph.SR},
       adsurl = {https://ui.adsabs.harvard.edu/abs/2024A&A...690L..15P}
}

@ARTICLE{Pietrow2024_flares,
       author = {{Pietrow}, A.~G.~M. and {Druett}, M.~K. and {Singh}, V.},
        title = "{Spectral variations within solar flare ribbons}",
      journal = {\aap},
         year = 2024,
        month = may,
       volume = {685},
          eid = {A137},
        pages = {A137},
          doi = {10.1051/0004-6361/202348839},
archivePrefix = {arXiv},
       eprint = {2402.10611},
 primaryClass = {astro-ph.SR},
       adsurl = {https://ui.adsabs.harvard.edu/abs/2024A&A...685A.137P}
}

@ARTICLE{Ruan2018,
       author = {{Ruan}, Guiping and {Schmieder}, Brigitte and {Mein}, Pierre and {Mein}, Nicole and {Labrosse}, Nicolas and {Gun{\'a}r}, Stanislav and {Chen}, Yao},
        title = "{On the Dynamic Nature of a Quiescent Prominence Observed by IRIS and MSDP Spectrographs}",
      journal = {\apj},
         year = 2018,
        month = oct,
       volume = {865},
       number = {2},
          eid = {123},
        pages = {123},
          doi = {10.3847/1538-4357/aada08},
       adsurl = {https://ui.adsabs.harvard.edu/abs/2018ApJ...865..123R}
}

@ARTICLE{Sahin2022,
       author = {{{\c{S}}ahin}, Seray and {Antolin}, Patrick},
        title = "{Prevalence of Thermal Nonequilibrium over an Active Region}",
      journal = {\apjl},
         year = 2022,
        month = jun,
       volume = {931},
       number = {2},
          eid = {L27},
        pages = {L27},
          doi = {10.3847/2041-8213/ac6fe9},
archivePrefix = {arXiv},
       eprint = {2205.10794},
 primaryClass = {astro-ph.SR},
       adsurl = {https://ui.adsabs.harvard.edu/abs/2022ApJ...931L..27S}
}

@ARTICLE{Sahin2023,
       author = {{{\c{S}}ahin}, Seray and {Antolin}, Patrick and {Froment}, Clara and {Schad}, Thomas A.},
        title = "{Spatial and Temporal Analysis of Quiescent Coronal Rain over an Active Region}",
      journal = {\apj},
         year = 2023,
        month = jun,
       volume = {950},
       number = {2},
          eid = {171},
        pages = {171},
          doi = {10.3847/1538-4357/acd44b},
archivePrefix = {arXiv},
       eprint = {2305.08775},
 primaryClass = {astro-ph.SR},
       adsurl = {https://ui.adsabs.harvard.edu/abs/2023ApJ...950..171S}
}

@ARTICLE{Sahin2024,
       author = {{{\c{S}}ahin}, Seray and {Antolin}, Patrick},
        title = "{From Chromospheric Evaporation to Coronal Rain: An Investigation of the Mass and Energy Cycle of a Flare}",
      journal = {\apj},
         year = 2024,
        month = aug,
       volume = {970},
       number = {2},
          eid = {106},
        pages = {106},
          doi = {10.3847/1538-4357/ad4ed9},
archivePrefix = {arXiv},
       eprint = {2406.02280},
 primaryClass = {astro-ph.SR},
       adsurl = {https://ui.adsabs.harvard.edu/abs/2024ApJ...970..106S}
}

@ARTICLE{Schwartz2024,
       author = {{Schwartz}, P. and {Gun{\'a}r}, S. and {Koza}, J. and {Heinzel}, P.},
        title = "{The diversity of spectral shapes of hydrogen Lyman lines and Mg II lines in a quiescent prominence}",
      journal = {\aap},
         year = 2024,
        month = apr,
       volume = {684},
          eid = {A197},
        pages = {A197},
          doi = {10.1051/0004-6361/202346251},
archivePrefix = {arXiv},
       eprint = {2401.09992},
 primaryClass = {astro-ph.SR},
       adsurl = {https://ui.adsabs.harvard.edu/abs/2024A&A...684A.197S}
}

@ARTICLE{Shibata_Magara2011,
       author = {{Shibata}, Kazunari and {Magara}, Tetsuya},
        title = "{Solar Flares: Magnetohydrodynamic Processes}",
      journal = {Living Reviews in Solar Physics},
         year = 2011,
        month = dec,
       volume = {8},
       number = {1},
          eid = {6},
        pages = {6},
          doi = {10.12942/lrsp-2011-6},
       adsurl = {https://ui.adsabs.harvard.edu/abs/2011LRSP....8....6S}
}

@ARTICLE{Snow2025,
       author = {{Snow}, B. and {Osborne}, C. and {Hillier}, A.~S.},
        title = "{Observational signatures of mixing-induced cooling in the Kelvin-Helmholtz instability}",
      journal = {\mnras},
         year = 2025,
        month = feb,
       volume = {537},
       number = {2},
        pages = {1904-1911},
          doi = {10.1093/mnras/staf144},
archivePrefix = {arXiv},
       eprint = {2501.11324},
 primaryClass = {astro-ph.SR},
       adsurl = {https://ui.adsabs.harvard.edu/abs/2025MNRAS.537.1904S}
}

@PROCEEDINGS{Vial_Engvold2015,
        title = "{Solar Prominences}",
    booktitle = {Solar Prominences},
         year = 2015,
       editor = {{Vial}, Jean-Claude and {Engvold}, Oddbj{\o}rn},
       series = {Astrophysics and Space Science Library},
       volume = {415},
        month = jan,
          doi = {10.1007/978-3-319-10416-4},
       adsurl = {https://ui.adsabs.harvard.edu/abs/2015ASSL..415.....V}
}

@ARTICLE{Vial2016,
       author = {{Vial}, Jean-Claude and {Pelouze}, Gabriel and {Heinzel}, Petr and {Kleint}, Lucia and {Anzer}, Ulrich},
        title = "{Observed IRIS Profiles of the h and k Doublet of Mg II and Comparison with Profiles from Quiescent Prominence NLTE Models}",
      journal = {\solphys},
         year = 2016,
        month = jan,
       volume = {291},
       number = {1},
        pages = {67-87},
          doi = {10.1007/s11207-015-0820-y},
       adsurl = {https://ui.adsabs.harvard.edu/abs/2016SoPh..291...67V}
}

@ARTICLE{Vial2019,
       author = {{Vial}, J.-C. and {Zhang}, P. and {Buchlin}, {\'E}.},
        title = "{Some relationships between radiative and atmospheric quantities through 1D NLTE modeling of prominences in the Mg II lines}",
      journal = {\aap},
         year = 2019,
        month = apr,
       volume = {624},
          eid = {A56},
        pages = {A56},
          doi = {10.1051/0004-6361/201834249},
       adsurl = {https://ui.adsabs.harvard.edu/abs/2019A&A...624A..56V}
}

@ARTICLE{Xia2011,
       author = {{Xia}, C. and {Chen}, P.~F. and {Keppens}, R. and {van Marle}, A.~J.},
        title = "{Formation of Solar Filaments by Steady and Nonsteady Chromospheric Heating}",
      journal = {\apj},
         year = 2011,
        month = aug,
       volume = {737},
       number = {1},
          eid = {27},
        pages = {27},
          doi = {10.1088/0004-637X/737/1/27},
archivePrefix = {arXiv},
       eprint = {1106.0094},
 primaryClass = {astro-ph.SR},
       adsurl = {https://ui.adsabs.harvard.edu/abs/2011ApJ...737...27X}
}

@ARTICLE{Xia2018,
       author = {{Xia}, C. and {Teunissen}, J. and {El Mellah}, I. and {Chan{\'e}}, E. and {Keppens}, R.},
        title = "{MPI-AMRVAC 2.0 for Solar and Astrophysical Applications}",
      journal = {\apjs},
         year = 2018,
        month = feb,
       volume = {234},
       number = {2},
          eid = {30},
        pages = {30},
          doi = {10.3847/1538-4365/aaa6c8},
archivePrefix = {arXiv},
       eprint = {1710.06140},
 primaryClass = {astro-ph.SR},
       adsurl = {https://ui.adsabs.harvard.edu/abs/2018ApJS..234...30X}
}

@ARTICLE{Watanabe2012,
       author = {{Watanabe}, K. and {Masuda}, S. and {Segawa}, T.},
        title = "{Hinode Flare Catalogue}",
      journal = {\solphys},
         year = 2012,
        month = jul,
       volume = {279},
       number = {1},
        pages = {317-322},
          doi = {10.1007/s11207-012-9983-y},
       adsurl = {https://ui.adsabs.harvard.edu/abs/2012SoPh..279..317W}
}

@ARTICLE{Wenzhi2024,
       author = {{Ruan}, Wenzhi and {Keppens}, Rony and {Yan}, Limei and {Antolin}, Patrick},
        title = "{The Lorentz Force at Work: Multiphase Magnetohydrodynamics throughout a Flare Lifespan}",
      journal = {\apj},
         year = 2024,
        month = jun,
       volume = {967},
       number = {2},
          eid = {82},
        pages = {82},
          doi = {10.3847/1538-4357/ad3915},
archivePrefix = {arXiv},
       eprint = {2403.19204},
 primaryClass = {astro-ph.SR},
       adsurl = {https://ui.adsabs.harvard.edu/abs/2024ApJ...967...82R}
}

@ARTICLE{Withbroe_Noyes1977,
       author = {{Withbroe}, G.~L. and {Noyes}, R.~W.},
        title = "{Mass and energy flow in the solar chromosphere and corona.}",
      journal = {\araa},
         year = 1977,
        month = jan,
       volume = {15},
        pages = {363-387},
          doi = {10.1146/annurev.aa.15.090177.002051},
       adsurl = {https://ui.adsabs.harvard.edu/abs/1977ARA&A..15..363W}
}

@ARTICLE{Gunar2024,
       author = {{Gun{\'a}r}, S. and {Heinzel}, P.},
        title = "{High-precision spectral inversions: Determining what is important for the accurate definition of incident radiation boundary conditions}",
      journal = {\aap},
         year = 2024,
        month = jul,
       volume = {687},
          eid = {A231},
        pages = {A231},
          doi = {10.1051/0004-6361/202449551},
       adsurl = {https://ui.adsabs.harvard.edu/abs/2024A&A...687A.231G}
}

\end{document}